\documentclass[12pt]{article}
\usepackage{epsfig,amssymb}

\newcommand{\bq}{\begin{equation}}
\newcommand{\eq}{\end{equation}}
\newcommand{\bqa}{\begin{eqnarray}}
\newcommand{\eqa}{\end{eqnarray}}
\newcommand{\ben}{\begin{enumerate}}
\newcommand{\een}{\end{enumerate}}
\newcommand{\bc}{\begin{center}}
\newcommand{\ec}{\end{center}}
\newcommand{\bqb}{\begin{eqnarray*}}
\newcommand{\eqb}{\end{eqnarray*}}

\def\pr#1#2#3{ Phys. Rev. ${\bf{#1}}$ (#2) #3}
\def\prl#1#2#3{ Phys. Rev. Lett. ${\bf{#1}}$ (#2) #3}
\def\pl#1#2#3{ Phys. Lett. ${\bf{#1}}$ (#2) #3}
\def\prep#1#2#3{ Phys. Rep. ${\bf{#1}}$ (#2) #3}

\def\np#1#2#3{ Nucl. Phys. ${\bf{#1}}$ (#2) #3}
\def\zp#1#2#3{ Z. f. Phys. ${\bf{#1}}$ (#2) #3}
\def\epj#1#2#3{ Eur. Phys. J. ${\bf{#1}}$ (#2) #3}

\def\ie{{\it i.e.\/}}
\def\eg{{\it e.g.\/}}

\def\etal{{\it et.al.\/}}

\global\nulldelimiterspace = 0pt

\def\wtil#1{\widetilde{#1}}

\def\L{ {\cal L }}

\def\sw{s_W}
\def\cw{c_W}
\def\swd{s^2_W}

\def\mz{m_Z}
\def\mzd{m_Z^2}
\def\mf{m_f}

\def\lam{\lambda}
\def\Lam{\Lambda}

\def\t{\hat t}
\def\s{\hat s}

\begin{document}
\pagenumbering{arabic}
\thispagestyle{empty}
\def\thefootnote{\fnsymbol{footnote}}
\setcounter{footnote}{1}

\begin{flushright}
PM/99-39 \\
THES-TP 99/11 \\
hep-ph/9910395 \\
October 1999\\
corrected version
 \end{flushright}
\vspace{2cm}
\begin{center}
{\Large\bf Signatures of the anomalous $Z \gamma $ and
$ZZ$ production at the lepton and hadron Colliders.}\footnote{Partially
supported by the grants ERBFMRX-CT96-0090 and CRG 971470.}
 \vspace{1.5cm}  \\
{\large G.J. Gounaris$^a$, J. Layssac$^b$  and F.M. Renard$^b$}\\
\vspace{0.7cm}
$^a$Department of Theoretical Physics, Aristotle
University of Thessaloniki,\\
Gr-54006, Thessaloniki, Greece.\\
\vspace{0.2cm}
$^b$Physique
Math\'{e}matique et Th\'{e}orique,
UMR 5825\\
Universit\'{e} Montpellier II,
 F-34095 Montpellier Cedex 5.\\
\vspace{0.2cm}

\vspace*{1cm}

{\bf Abstract}
\end{center}

The possible form of New Physics (NP) interactions
affecting the $ZZZ$, $ZZ \gamma $ and $Z\gamma \gamma$
vertices, is critically examined.
Their signatures and the possibilities
to  study them, through  $ZZ$ and $Z\gamma $ production,
at the $e^-e^+$ Colliders LEP and LC and at the
hadronic Colliders Tevatron and LHC, are investigated. Experimental
limits obtained or expected on each coupling are collected.
A simple theoretical model based on virtual effects
due to some heavy fermions is used for acquiring
some guidance on the plausible forms of
these NP vertices. In such a case specific relations among
the various neutral couplings are predicted, which
can be experimentally tested and possibly used to constrain
the form of the responsible  NP structure.\par

\def\thefootnote{\arabic{footnote}}
\setcounter{footnote}{0}
\clearpage

\section{Introduction}

During the last two decades an intense activity has taken place
about the possible existence of anomalous gauge boson couplings (i.e.
non standard contributions). The
general form of the 3-boson couplings was written, in a
model-independent way, in terms of a set of seven independent
Lorentz and $U(1)_{em}$ invariant operators\cite{GG-TGC,
Hagiwara}. This general description has been applied to both
charged ($\gamma W W$, $ZWW$) and neutral ($\gamma \gamma Z$,
$\gamma ZZ$, $ZZZ$) sectors \cite{Hagiwara}.\par

More recently, anomalous gauge boson couplings
were considered in the framework of the Effective
Lagrangians \cite{Georgi}.
Here, the basic assumption is that, beyond SM, there
exist  new physics (NP) dynamics whose degrees of freedom are so
heavy (of mass scale $\Lambda$) that they cannot be produced at
present or near future colliders. The only observable effects
should then be anomalous interactions of usual SM particles. Under
these conditions, by integrating out these heavy NP states,
the observable effects can be described by an Effective Lagrangian
constructed in terms of operators involving only SM fields
\cite{Hagiwara, RG-effective}. So long as $\Lambda$ is much larger than
the actually observable  energy range,
these operators are dominated by those with the lowest possible
dimension. Each operator should be hermitian, multiplied by a
constant coupling; while contributions from higher dimensional operators
should be suppressed by powers of $s/\Lam^2$.\par

The set of anomalous couplings can be classified and restricted
using  symmetry requirements and constraints on the highest
allowed dimensionality.  This procedure has been fruitfully
applied to various sectors of the SM \cite{Buchmuller}. Thus, it
has allowed to describe anomalous properties of
several processes, like 4-fermion, 2 fermion-2 boson, 3-boson, 4-boson
interactions; where the fermions are leptons or  light
or heavy quarks, while the bosons are $\gamma$, W, Z and Higgs.\par

The charged 3-boson sector has been explored in great detail with
this method, both theoretically and experimentally \cite{Wudka, work}.
The general
form with 7 types of couplings (4 CP-conserving and 3
CP-violating for the photon and separately for the  $Z$  also),
was shown to be reduced  to only 5 independent couplings (3
CP-conserving and 2 CP-violating) if one restricts to $dim=6$
$SU(2)\times U(1)$ gauge invariant operators in the linear
representation \cite{Hagiwara2}. While, in the non-linear
representation case (where no light Higgs boson exists),
one finds that 4 independent CP conserving
and 3 CP violating $SU(2)\times U(1)$ gauge
invariant operators contribute to triple gauge couplings,
at the level of $d_{chiral}=4$  \cite{Yuan}. Various other assumptions
can also reduce the number of independent couplings \cite{Schi}.\par

Experimental constraints have already been established through $W^+W^-$
production at LEP2 and $W\gamma$, $WZ$ production at the TEVATRON,
\cite{LEPexp, CDF, D0}.
Relations between the coupling constants and the effective NP scale
$\Lambda$ have also been established through unitarity
relations; which allow to translate the upper limits on
these couplings into lower limits for the effective scale
$\Lam$  \cite{unit}. Using this framework, a comparison
of the experimental results already obtained or expected at
future colliders in the various processes, should allow to establish
interesting constraints on the possible structure of the NP
interactions. At least it should show what is the  SM sector that
NP may affect, and what symmetry property it may preserve.\par

Our first aim in this paper is to explore if similar information
could be obtained
in the neutral 3-boson sector. Up to now, this sector has received less
attention than the charged one. Probably this is because
charged boson couplings already received tree level Standard Model
(SM) contributions, whereas the neutral ones do not; so that they
may  be considered as purely "anomalous". The situation in it  is
less simple for several reasons. To the general Lorentz and $U(1)_{em}$
invariance requirements, one should add the constraints due to Bose
statistics, as there are always at least two identical particles.
This forbids  $ZZZ$, $ZZ\gamma $
or $Z\gamma \gamma$ interactions vertices when all particles are
on-shell \cite{GG-TGC}. The appearance of such vertices is only
possible if at least one of the gauge bosons involved is off
shell. The first discussions about these couplings were
given in \cite{FFB}. The most general allowed form
involves only 2 independent
couplings for each of the $VZZ$ vertices ($V=\gamma,~Z$; one
CP-conserving and one CP-violating) and 4 independent couplings
for each of the $VZ\gamma$ vertices ($V=\gamma,~Z$; two
CP-conserving and two CP-violating). There is a priori no relation
between these various couplings. Explicit
expressions for these vertices were written in
\cite{Hagiwara} and have then be widely used. However we noticed that
a factor $i$ was omitted in the set of $VZ\gamma$ vertices.
This factor $i$ is absolutely necessary
in order the related effective
New Physics (NP) Lagrangian to be hermitian.\par

As in the charged 3-boson sector, this effective lagrangian may be
written  in  an $SU(2)\times U(1)$ invariant form.
The only difference is  that, while in the charged sector the
NP interactions maybe generated already at the level of
$dim=6$ operators\footnote{We assume here the linear scalar sector
representation}; in the neutral sector we need
operators of dimensions 8 or 10 in order for NP to be generated.
So, if we restrict to $dim=6$ operators,
no NP vertices  in the neutral 3-boson sector is allowed.
 Thus, if such interactions exist, it would indicate either
that some  higher  dimensional operators containing
neutral 3-boson vertices without appreciable admixture from
charged ones are somehow enhanced; or
that the NP scale is rather nearby, so that there is no
dimensional ordering on the size of the various operators.
But of course, in such a case direct production of the new degrees
of freedom may be observable.
 This fact should also arise
when one tries to write unitarity  constraints and relate the neutral
couplings to the effective NP scale defined as the energy at which
the various amplitudes saturate unitarity \cite{unit}.
To be more precise we take one example of NP structure
due to the one loop virtual effects of heavy fermions, and we
discuss the corresponding pattern of anomalous couplings that are
generated. It is found then the strength of these couplings may be
enhanced compared to what the dimensionality of the related
operators would had led us to expect. Moreover
relations among the various couplings are obtained in such models.
It will be very interesting to see what constraints
the experimental measurements will put on these couplings;
 i.e. to see how they compare to the above theoretical
pattern in the neutral and in the charged sectors.\par

Thus, our motivation for  reconsidering
the various $ZZ$ and $Z\gamma$ production processes at LEP2, LC,
TEVATRON and LHC, is to see how they react
to the presence of each of the anomalous
couplings. In the next Section 2 we explicitly write the correct
neutral 3-boson vertices and the Effective Lagrangian from which they
derive. A toy model for the generation of such couplings is
also presented.
We then give  the corresponding NP contributions to the
helicity amplitudes for the $f\bar f\to
ZZ,~Z\gamma$ processes. Our conventions are fully defined
by the expressions for the SM parts of the amplitudes that we
give  in Appendix A.  The expressions of the observables (cross
sections and asymmetries) at the various colliders are given in
Appendix B. In Section 3 we give explicit illustrations showing how the
observables react to each of the anomalous couplings, in particular the
interference patterns for the case of CP-conserving couplings. We
emphasize the special role that  longitudinal polarization would play at
the LC Collider, for disentangling photon and $Z$ anomalous couplings.
We also devote a special attention to the way these anomalous effects
would be analyzed at hadron Colliders and the respective merits of
transverse momentum, invariant mass and c.m. scattering angle
distributions. Finally we summarize our observations and suggestions in
Sect.4 .

\section{Description of anomalous neutral boson couplings}

\begin{figure}[p]
\vspace*{-4cm}
\[
\epsfig{file=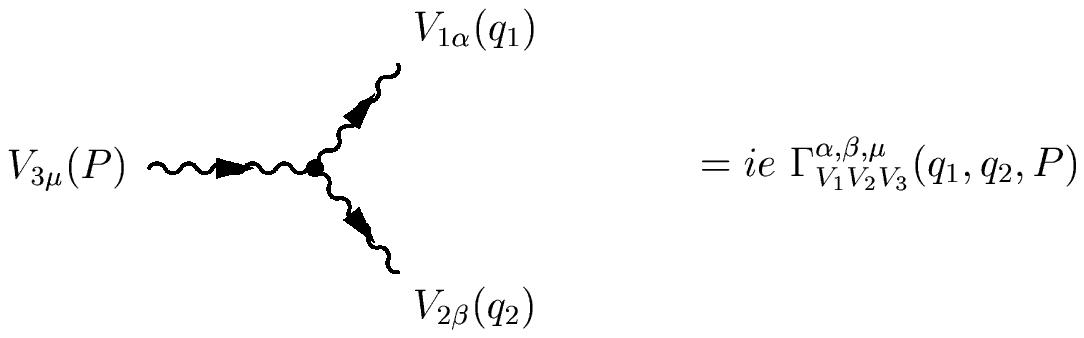,height=3cm,width=12cm}
\]
\vspace*{0.5cm}
\caption[1]{Feynman rule for the general $V_1V_2V_3$ vertex.}
\label{Feyn-fig}
\end{figure}

Assuming only
Lorentz and $U(1)_{em}$ gauge invariance as well as Bose statistics,
the most general form of the $V_1V_2V_3$ vertex function
defined in Fig.\ref{Feyn-fig}, where $V_1,V_2$ are on shell
 neutral gauge bosons, while $(V_3=Z~,~\gamma) $  is in general
off-shell but always coupled to a conserved current, has been
given in\footnote{We define $\epsilon^{0123}=+1$.}  \cite{Hagiwara}
\bqa
\Gamma^{\alpha \beta \mu}_{ZZ V} (q_1, q_2, P)
&=& \frac{i (P^2-m_V^2)}{\mzd}
\left [ f_4^V (P^\alpha g^{\mu \beta}+P^\beta g^{\mu \alpha})
-f_5^V \epsilon^{\mu \alpha \beta \rho}(q_1-q_2)_\rho \right ]
~, \label{fZZ} \\
\Gamma^{\alpha \beta \mu}_{Z\gamma V} (q_1, q_2, P)
&=& \frac{ i (P^2-m_V^2)}{\mzd}
\Bigg \{ h_1^V (q_2^\mu g^{\alpha \beta}-q_2^\alpha g^{\mu \beta} )
+ \frac{h_2^V}{\mzd} P^\alpha [ (Pq_2) g^{\mu \beta}- q_2^\mu P^\beta ]
\nonumber \\
&-& h_3^V \epsilon^{\mu \alpha \beta \rho} q_{2\rho}
~-~\frac{h_4^V}{\mzd} P^\alpha \epsilon^{\mu \beta \rho
\sigma}P_\rho q_{2\sigma} \Bigg \}~ . \label{hZgamma}
\eqa

Compared to \cite{Hagiwara}, we have introduced in (\ref{hZgamma})
an additional factor $i$ in order for the related effective
New Physics (NP) Lagrangian to be hermitian. Of course,
the choice of the sign of this factor is
a convention.\par

The effective Lagrangian generating the vertices (\ref{fZZ},
\ref{hZgamma}) is\footnote{Some specific terms of this Lagrangian
have been considered in \cite{Boudjema}.}
\bqa
\L_{NP} &=& \frac{e}{\mzd} \Bigg [
-[f_4^\gamma (\partial_\mu F^{\mu \beta})+
f_4^Z (\partial_\mu Z^{\mu \beta}) ] Z_\alpha
( \partial^\alpha Z_\beta)+
[f_5^\gamma (\partial^\sigma F_{\sigma \mu})+
f_5^Z (\partial^\sigma Z_{\sigma \mu}) ] \wtil{Z}^{\mu \beta} Z_\beta
\nonumber \\
&-&  [h_1^\gamma (\partial^\sigma F_{\sigma \mu})
+h_1^Z (\partial^\sigma Z_{\sigma \mu})] Z_\beta F^{\mu \beta}
-[h_3^\gamma  (\partial_\sigma F^{\sigma \rho})
+ h_3^Z  (\partial_\sigma Z^{\sigma \rho})] Z^\alpha
 \wtil{F}_{\rho \alpha}
\nonumber \\
&- & \left \{\frac{h_2^\gamma}{\mzd} [\partial_\alpha \partial_\beta
\partial^\rho F_{\rho \mu} ]
+\frac{h_2^Z}{\mzd} [\partial_\alpha \partial_\beta
(\square +\mzd) Z_\mu] \right \} Z^\alpha F^{\mu \beta}
\nonumber \\
&+& \left \{
\frac{h_4^\gamma}{2\mzd}[\square \partial^\sigma
F^{\rho \alpha}] +
\frac{h_4^Z}{2 \mzd} [(\square +\mzd) \partial^\sigma
Z^{\rho \alpha}] \right \} Z_\sigma \wtil{F}_{\rho \alpha }
 \Bigg ] ~ , \label{LNP}
\eqa
where $\wtil{Z}_{\mu \nu}=1/2 \epsilon_{\mu \nu \rho \sigma}Z^{\rho
\sigma}$ with
$Z_{\mu\nu}=\partial_\mu Z_\nu -\partial_\nu Z_\mu$ and similarely for
the photon tensor $F_{\mu\nu}$.
The couplings
$f_4^V, ~h_1^V,~ h_2^V$ violate CP invariance;
while $f_5^V, ~h_3^V ,~ h_4^V$ respect it. \par

The use of the equations of motion for the photon and Z-fields
implies that the replacements
\bqa
\partial^\mu F_{\mu \nu} & \Longrightarrow &
eQ_{f} \bar f \gamma_\nu f  ~ , \label{Jgamma}\\
\partial^\mu Z_{\mu \nu}+ \mzd Z_\nu& \Longrightarrow &
e \bar f \left  (g^Z_L \gamma_\nu \frac{(1-\gamma_5)}{2}  ~+ ~
g^Z_R \gamma_\nu \frac{(1+\gamma_5)}{2} \right ) f ~ ,
\label{JZ1}\\
(\square + \mzd) Z_\nu& \Longrightarrow &
e \bar f \left  (g^Z_L \gamma_\nu \frac{(1-\gamma_5)}{2}  ~+ ~
g^Z_R \gamma_\nu \frac{(1+\gamma_5)}{2} \right ) f ~ ,
\label{JZ2}
\eqa
may be done in the first factor of each term in (\ref{LNP}),
where $f$ is  any fermion  with couplings defined in (\ref{gZ}).
Thus, the effective Lagrangian in (\ref{LNP}) is essentially
equivalent to a set of  contact $f\bar f ZZ$ and $f \bar f Z\gamma $
interactions. \par

Of course, the computation of  the NP
scattering amplitudes for
$f\bar f \to ZZ$ and $f\bar f \to Z\gamma$,
 either by using these contact interactions,  or
working directly with (\ref{LNP}), gives the same results.
They are given below, and should be added to the SM ones, which are
 due to fermion ($f$) exchange in the $t$-channel. These
SM helicity amplitudes appear in Appendix A and serve to
define our notations and conventions.\par

In $f \bar f \to ZZ$, the only non-vanishing NP helicity amplitudes
induced by (\ref{fZZ}) are those where one $Z$ is
transverse ($\tau_1\equiv\tau=\pm 1$) and the other longitudinal
($\tau_2=0$). In this case  we have
\bqa
&& F^\lambda_{\tau 0}( f \bar f \to  ZZ~;~NP)=
F^{\lambda *}_{0, - \tau}( f \bar f \to  ZZ~;~NP)
\nonumber \\
&=& \frac{e^2 \s^{3/2}\beta}{\mz^3 2 \sqrt{2}}
(1+\lam\tau \cos\vartheta^*)\left [
i (f_4^\gamma Q_f +f_4^Z g_\lam^Z)-
(f_5^\gamma Q_f +f_5^Z g_\lam^Z)\beta \tau \right ] ~,
\label{FffZZ-NP}
\eqa
where the same definitions as in (\ref{FffZZ}) are used.\par

Correspondingly for the NP contribution to
$f \bar f \to Z \gamma $, compare (\ref{FffZg})
\bqa
F^\lambda_{\tau_1 \tau_2}( f \bar f \to  Z \gamma ~;~ NP)
&= & -~\frac{ e^2 (\s-\mzd)\lam }{4 \mzd} \sin\vartheta^*
\Big [ i (h_1^\gamma Q_f +h_1^Z g_\lam^Z) (1 +\tau_1\tau_2)
\nonumber \\
&& -(\tau_1+\tau_2)(h_3^\gamma Q_f +h_3^Z g_\lam^Z)
\Big ] ~ \mbox{ for } ~ \tau_1 \tau_2 \neq 0 ~~ , \label{FffZg-NP1}
\eqa
\bqa
&& F^\lambda_{0 \tau_2}( f \bar f \to  Z \gamma ~;~ NP)
=  -~ \frac{ e^2 \sqrt{\s} (\s-\mzd)}{\mz^3 2 \sqrt{2}}
(1-\lam \tau_2 \cos\vartheta^*) \Big [
-i (h_1^\gamma Q_f +h_1^Z g_\lam^Z)
\nonumber \\
&& +i (h_2^\gamma Q_f +h_2^Z g_\lam^Z)\frac{(\s-\mzd)}{2\mzd}
 + \tau_2  (h_3^\gamma Q_f +h_3^Z g_\lam^Z)
-\tau_2 \frac{(\s-\mzd)}{2\mzd} (h_4^\gamma Q_f +h_4^Z g_\lam^Z)
\Big ] ~ , \label{FffZg-NP2}
\eqa
where of course $\tau_2 =\pm 1$.

\newpage
\noindent
{\bf A toy model: heavy fermion contributions at one loop.}\par
In order to give at least one illustration of how such anomalous
couplings can be generated, we consider the virtual effects of heavy
particles at one loop (triangle diagrams with $\gamma$ and $Z$ external
legs), using standard gauge boson couplings.
We first observe that heavy scalar particles cannot generate
such neutral self-couplings.
Heavy fermions can generate $f^V_5$ and $h^V_3$ couplings
($V=\gamma,~Z$). No CP-violating couplings ($f^V_4$, $h^V_{1,2})$
and no $h^V_4$ coupling are generated at this level. Higher order
effects are needed to get them, see \cite{model} for a detailed
discussion.\par

These results suggest that, indeed, the dominant anomalous couplings
may be $f^V_5$ and $h^V_3$. In fact at one loop, the results
of the computation in \cite{model} for a
heavy fermion $F$ interacting with $Z$ and $\gamma$
as
\bq \L = -eQ_FA^\mu\bar F \gamma_\mu F
-{e\over2s_Wc_W}  Z^\mu \bar F \left  ( \gamma_\mu g_{VF}
- \gamma_\mu \gamma_5 g_{AF} \right ) F  ~~ , \label{VheavyF}
\eq
give
\bqa
&& h^Z_3=-f^{\gamma}_5=
-N_f{e^2 Q_F g_{VF} g_{AF}\over96\pi^2s^2_Wc^2_W}
\left ({m^2_Z\over M^2_F}\right ) ~~, \label{h3Zmodel} \\
&& h^{\gamma}_3
=-N_F{e^2 Q^2_F g_{AF}\over48\pi^2s_Wc_W}
\left ({m^2_Z\over M^2_F}\right ) ~ ~ , \label{h3gmodel}\\
&& f^{Z}_5 =N_f\frac{e^2 g_{AF} (5 g_{VF}^2 +g^2_{AF})}
{960\pi^2s^3_Wc^3_W} \left({m^2_Z\over M^2_F}\right ) ~~
, \label{f5Zmodel}\\
&& h^Z_4=h^{\gamma}_4=0  ~ ~ , \label{h4Zmodel}
\eqa
\noindent
where $Q_F$ is the $F$ electric charge,  and $g_{VF}$, $g_{AF}$ are
defined in (\ref{VheavyF}).  $N_F$ is a
(colour, hypercolour) counting
factor which may possibly include enhancement effects due to a strongly
interacting sector, while $M_F$ is the $F$ mass.\par

In general there is no relation to be expected
between $f^V_i$ and $h^V_i$ couplings.  Note though from
(\ref{h3Zmodel}), that in the above model the remarkable relation
\bq
h^Z_3=-f^{\gamma}_5 ~~ \label{h3Z-f5g}
\eq
should hold, which is independent of the fermion couplings.
Another striking result is that  there are no  $h^Z_4$ or $h^{\gamma}_4$
couplings   in such a model \cite{model}. We also remark that such a
model would also generate anapole $ZWW$ and $\gamma WW$
couplings, when the heavy fermion is integrated out
 at the 1-loop level. \par

Of course, a complete family of exactly degenerate heavy fermions
(leptons and quarks with the SM structure) would lead to the vanishing
of all the NP couplings in (\ref{h3Zmodel}-\ref{f5Zmodel}).
Because, in this case the combination of the heavy fermion
contributions is the same as in the (mass independent) cancellation
in the triangle anomaly. This is the
unbroken $SU(2)\times U(1)$ situation.\par

If instead,  one introduces a mass splitting of electroweak size
(i.e. $\simeq m^2_Z$ )  among the multiplets;
like \eg\@ between the heavy lepton and  quark doublets;  then
the resulting couplings are of the order ${m^4_Z/ M^4_F}$,
which means that they are suppressed by an extra
power  $\mzd/M_F^2$, as compared to what appears in
(\ref{h3Zmodel}-\ref{f5Zmodel}). This
case is referred to as a spontaneous broken $SU(2)\times U(1)$
situation in \cite{model}. \par

Finally, if a single (or a  doublet of a)  heavy fermion is much lighter
than all the other fermions in the family, then the couplings are
as appearing in (\ref{h3Zmodel}-\ref{f5Zmodel}); \ie\@
just proportional to $({m^2_Z/ M^2_F})$.
This is obviously the most favorable situation
for their observability,   and would essentially
mean that $SU(2)\times U(1)$ is strongly broken in the NP sector. \par

A final important warning concerning the magnitude
of the above couplings must be made.
Keeping only standard gauge couplings, the factor $\alpha/4\pi$ which
naturally arises in the one loop computations, predicts anomalous
couplings of the order
of $10^{-3}$ for $M_F$ in the $100~GeV$ range. So without a strong
enhancement factor there is little hope of observability, except
with the very high luminosities expected for the LC collider as we will
see in the next Section.

\section{Application to $ZZ$ and $Z\gamma$ production processes}

In this Section we examine how the presence of any of the
aforementioned anomalous couplings reflects in $ZZ$ and $Z\gamma$
production at present and future $e^+e^-$ and hadron colliders.
The corresponding differential cross sections are given in Appendix
B. They are expressed in terms of helicity amplitudes for the
basic $f\bar f\to ZZ$ and $f\bar f\to Z\gamma$ processes. \par

As expected, the CP-conserving couplings always
lead to real amplitudes interfering with the SM ones;
so that  the various observables
are linearly  sensitive to these NP terms. On the contrary,  the
CP-violating couplings always  lead to purely imaginary amplitudes
that do not interfere with the SM
ones\footnote{A small interference could only arise for $Z\gamma$
production at energies  rather close to the $Z$-pole, where Z-width
effects may be non-negligible.}. Thus the CP violating
observables  depend only quadratically on the NP couplings, and their
sensitivity limits are accordingly reduced.\par

Another feature is related to the dimension ($dim=6$ and
$dim=8$) of the couplings in the Lorentz and $U(1)_{em}$ invariant
expression   (\ref{LNP}).
Obviously the $dim=8$ couplings $h^V_2$ and $h^V_4$,
associated to terms growing with one more power of $\hat s$, will
be more easily constrained than the $dim=6$ ones, thus affording
a better sensitivity limit.\par
\vspace{0.5cm}

\noindent
{\bf Form factors:}\par
 Especially  at hadron colliders,
it has become rather usual to analyze the
NP sensitivity limits, by multiplying
the basic constant anomalous couplings defined in Sect. 2,
by "form factors" \cite{Baur, Wudka}.
The reason for this procedure is the
following. For a given value of these basic couplings, (for
example chosen in order to give a visible effect at an
intermediate $\sqrt{\hat s}$ energy), the departure from the  SM
prediction grows rapidly when $\hat s$ increases, and may
even reach an unreasonable (unitarity violating) size. In order to
cure this behaviour, form factors decreasing with $\hat s$
with an arbitrary scale
(denoted below as  $\Lambda_{FF}$), are introduced.
The form factor usually
used is $(1+\hat s/\Lambda_{FF}^2)^{-n}$,
with n=3 for $f^V_{4,5}$, $h^V_{1,3}$ and n=4 for $h^V_{2,4}$ \cite{D0}.
In our illustrations we shall neglect the form factor at LEP2;
but, for comparison with previous works,  we shall keep it for LC where
we take $\Lambda_{FF}=1~TeV$, as well as for the
TEVATRON for which we take  $\Lambda_{FF}=0.75~TeV$, and LHC  for
which  $\Lambda_{FF}=3~TeV$ is used.\par

When one analyses experimental results at a given
$\sqrt{\hat s}$, it is not of particular  importance
whether one chooses to use or not to use
this procedure; as one can unambiguously translate the limits
obtained with form factors, to those
reached without them.
However at a hadron collider  where the  limits often arise
  from an integration over a large range of
$\hat s$, no simple correspondence is possible.\par

In fact, the use of form factors is somewhat in contradiction
with the basic assumption ($\Lambda \gg \sqrt{\hat s}$), that allows to
work with Effective Lagrangians keeping only the lowest dimensions. The
additional $\hat s$ dependence brought in by the form factor, would
correspond to the presence of higher dimensional operators with a
specific form. Therefore, we would prefer  a treatment
where  no form factors are used, and   one instead tries to stay within
the basic assumptions; \ie\@ to keep working within the range
$\Lambda \gg \sqrt{\hat s}$ and far from the unitarity limit, by
considering sufficiently small values for the anomalous couplings
for each $\sqrt{\hat s}$ domain.
We shall come back to this point with some new proposal at the end of
this Section.\par

\begin{figure}[p]
\vspace*{-4cm}
\[
\epsfig{file=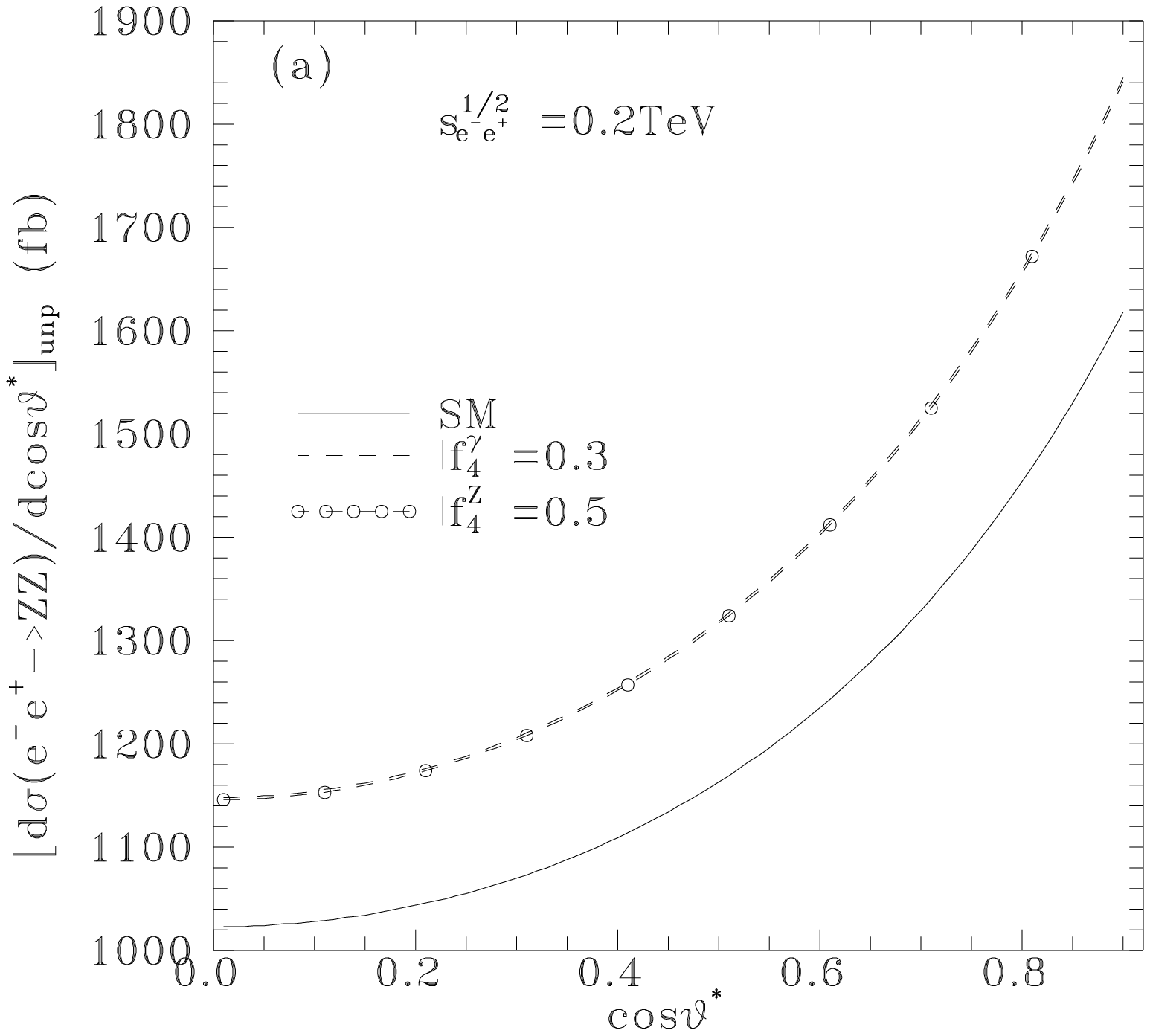,height=7.5cm}\hspace{0.5cm}
\epsfig{file=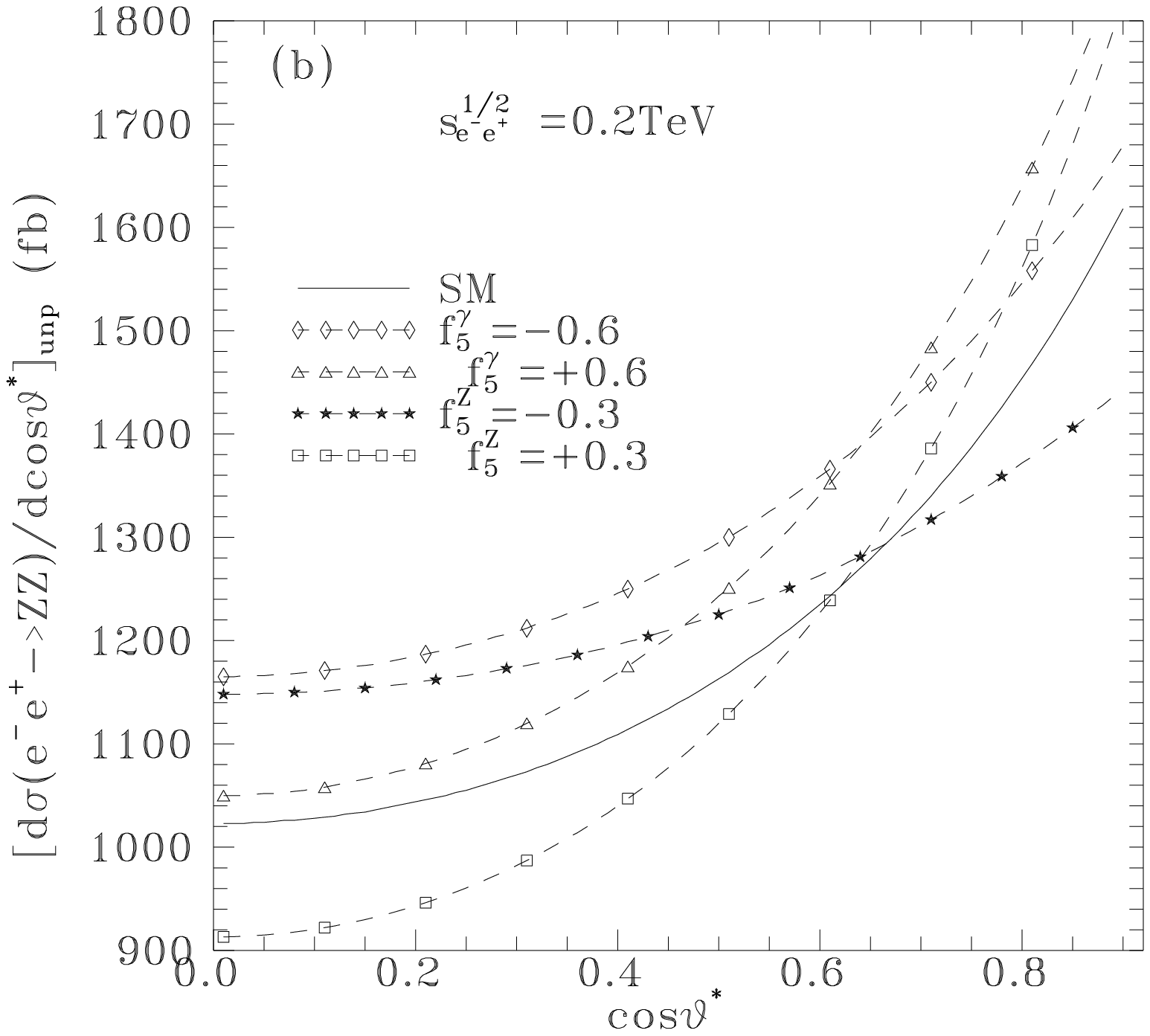,height=7.5cm}
\]
\vspace*{0.5cm}
\caption[1]{Standard and anomalous contributions to the
unpolarized $e^-e^+ \to ZZ $ cross section at  LEP.}
\label{LEP-fig-ZZ}
\end{figure}

\vspace{0.5cm} \noindent {\bf Application to LEP2 at 200 GeV}\\
The results for \underline{$e^+e^-\to ZZ$} are shown in
Fig.\ref{LEP-fig-ZZ}a,b. As expected, the non interfering
CP-violating couplings always produce an increase of the cross
section; whereas CP-conserving ones produce typical interference
patterns with the SM contribution.\par

The final sensitivity will depend on the integrated luminosity
assumed here to be $150~ pb^{-1}$; and on the angular cuts and
selection of $Z$ decay modes needed for its identification, which
should reduce the number of events by roughly a
factor 2. In Fig.2a,b we illustrate the additive effects of the
CP-violating couplings with $|f^{\gamma}_4|=0.3$ and
$|f^Z_4|=0.5$, and the interference patterns of the CP-conserving ones
with $f^{\gamma}_5=\pm0.6$ and $f^{Z}_5=\pm0.3$. With the expected
number of events these values roughly correspond to one standard
deviation from SM predictions. This may be compared with recent
results obtained at 189 GeV (see e.g. \cite{L3-ZZ}), in which
observability limits were given at 95 \% confidence level:

\bq
-1.9\leq f^Z_4\leq 1.9 \ , \ -5.0\leq f^Z_5\leq 4.5 \ , \
-1.1\leq f^{\gamma}_4\leq 1.2 \ ,
\ -3.0\leq f^{\gamma}_5\leq 2.9 ~.
\eq

Note that around 200 GeV we are just above the $ZZ$ threshold
where the beta factor (compare (\ref{dsig-ffZZ})) strongly affects
the cross section. Thus, in this region, the sensitivity to
the $f^{V}_5$ couplings strongly increase with the energy.

\begin{figure}[p]
\vspace*{-4cm}
\[
\epsfig{file=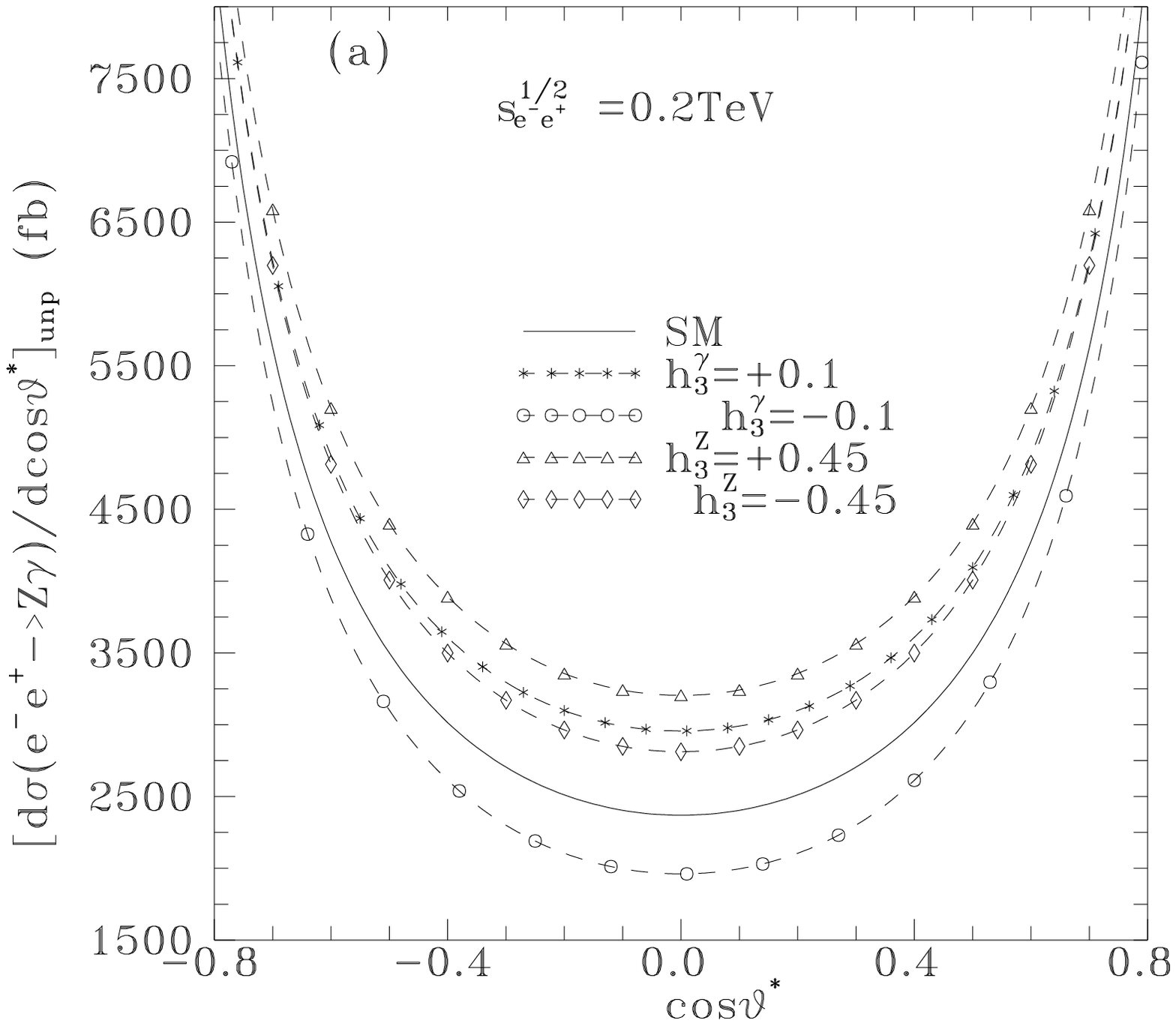,height=7.5cm}\hspace{0.5cm}
\epsfig{file=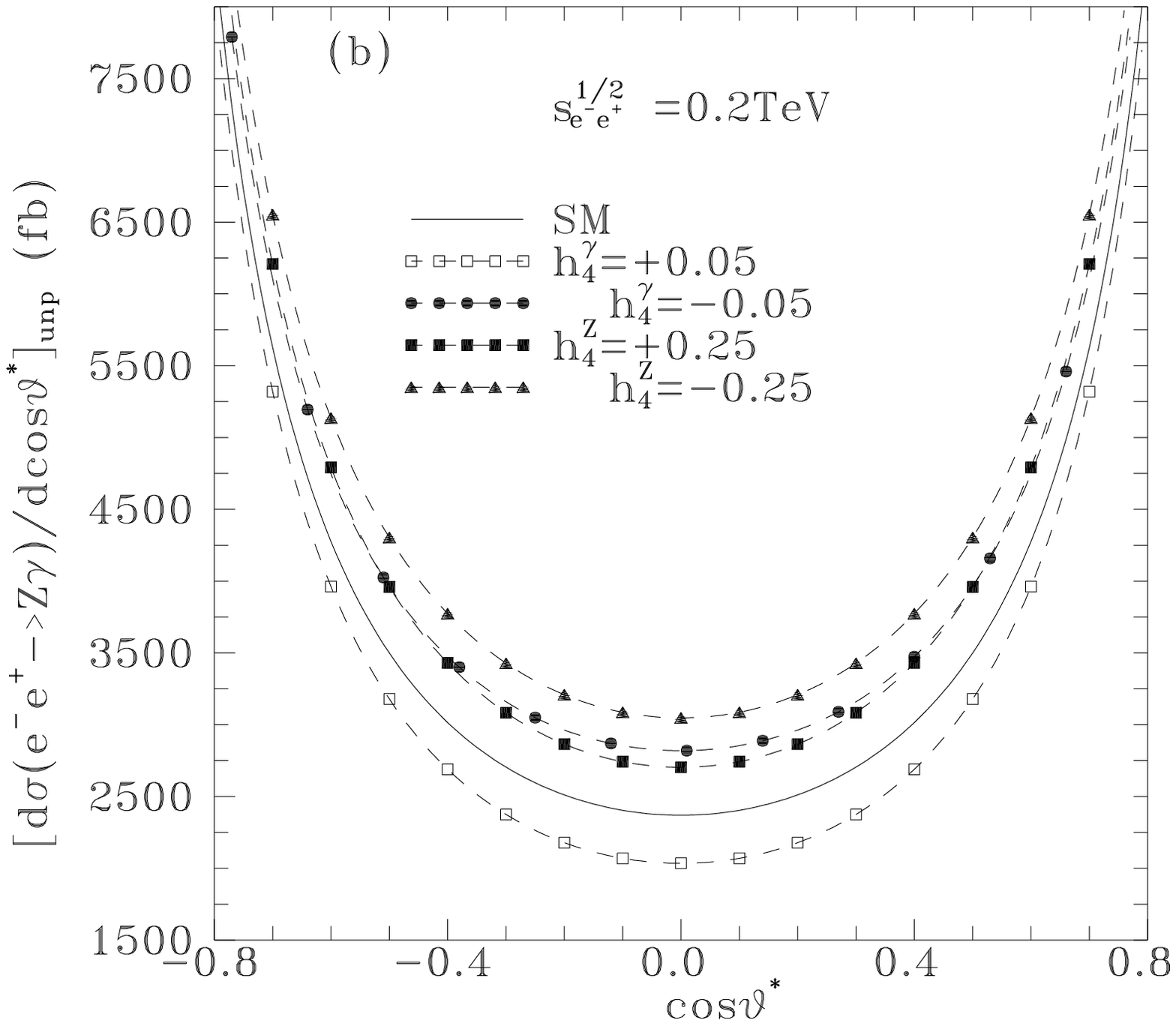,height=7.5cm}
\]
\vspace*{0.5cm}
\[
\epsfig{file=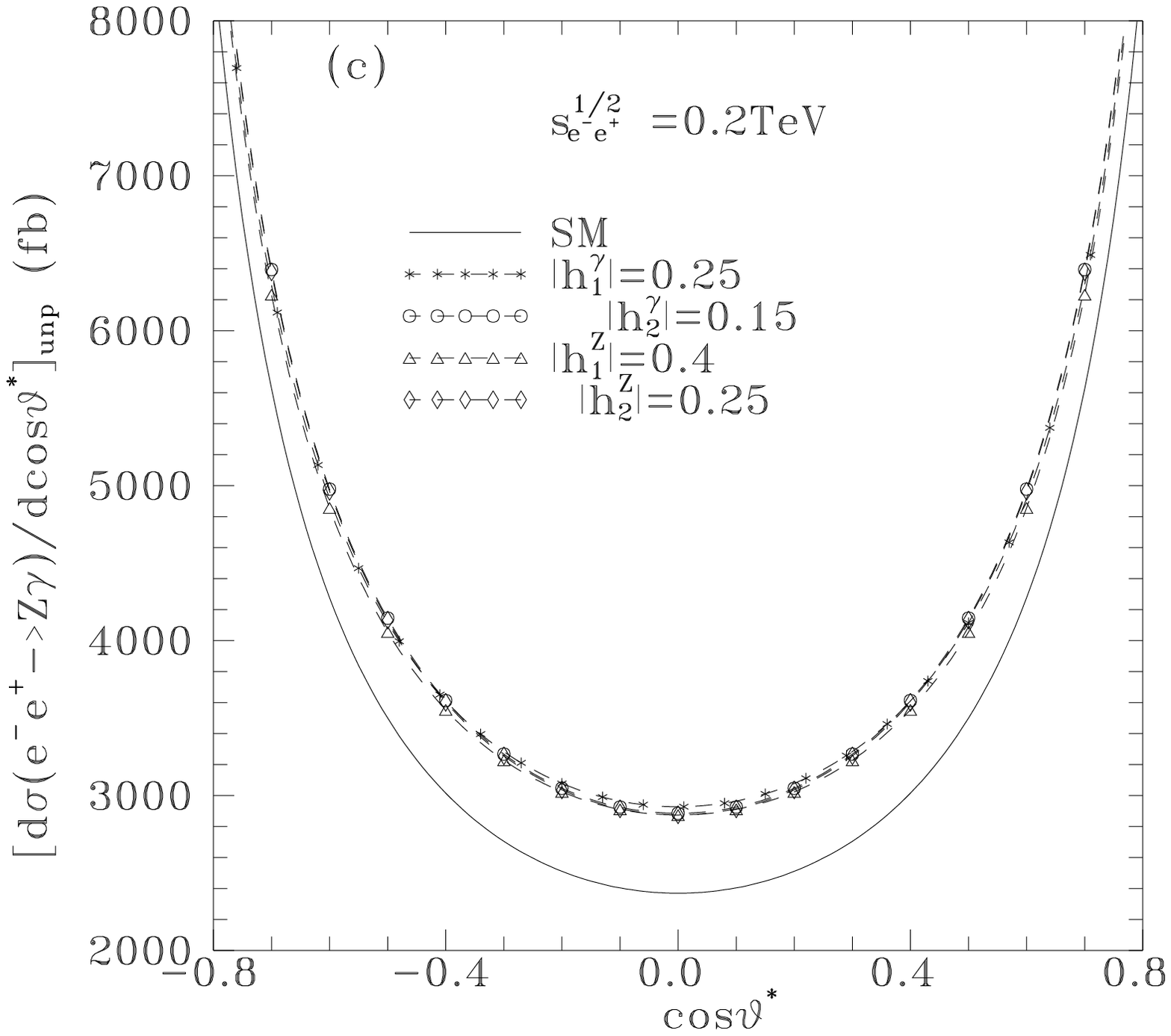,height=7.5cm}
\]
\vspace*{0.5cm}
\caption[1]{Standard and anomalous contributions to the unpolarized
$e^-e^+ \to Z \gamma$ cross sections at  LEP.}
\label{LEP-fig-Zg}
\end{figure}

In the case of \underline{$e^+e^-\to Z\gamma$},
the cross section is larger than the $ZZ$ one.
It is a factor 2 larger at large angles, and it has
a much larger forward peaking;
see Fig.\ref{LEP-fig-Zg}a,b,c. Since the detection of
the final photon should be
sufficient to characterize the process and all Z decay modes
may be used,  no reduction factor is probably needed.
This should lead to a number of events
an order of magnitude larger than for  $ZZ$. Consequently
the observability limits should be much better.
For the CP-violating  couplings we then  expect one standard deviation
effects like   $|h^{\gamma}_1|=0.1$, $|h^{Z}_1|=0.2$,
$|h^{\gamma}_2|=0.07$ ,$|h^{Z}_2|=0.12$.
Correspondingly, for the CP-conserving  couplings, we expect
asymmetrical one standard deviation effects of the form
$h^{\gamma}_3=\pm0.02$, $h^{Z}_3=\pm0.12$, $h^{\gamma}_4=\pm0.015$,
$h^{Z}_4=\pm0.09$.\par

The difference in the sensitivities to the  $\gamma$- and Z-couplings
can be simply understood as a consequence of the fact that
the exchanged photon has a pure vector electron  coupling $Q_e$,
whereas the exchanged Z has a weaker (by a factor $4s_Wc_W$)
and essentially
purely axial  coupling to the electron, so that the interference patterns
with the SM amplitude differ in size and in sign for each helicity
amplitude; the interplay of linear and quadratic contributions
generates further differences.

\vspace{0.5cm}
\noindent
{\bf Application to LC at 500 GeV}\\
At energies of $500~GeV$ and at large angles,
the cross section is weaker than at 200 GeV
by about a factor 10.
This should be largely compensated by the expected increase in
luminosity \cite{LC} (three orders of magnitude for TESLA);
which  leads to a number of
events larger by more than two orders of magnitude.
In addition,  the NP amplitude increases like $s$ (or even $s^2$),
producing at least an
additional order of magnitude in the sensitivity. So finally,
the statistical sensitivity to the above couplings should be
increased by \underline{more than two orders of magnitude}.\par

\begin{figure}[p]
\vspace*{-4cm}
\[
\epsfig{file=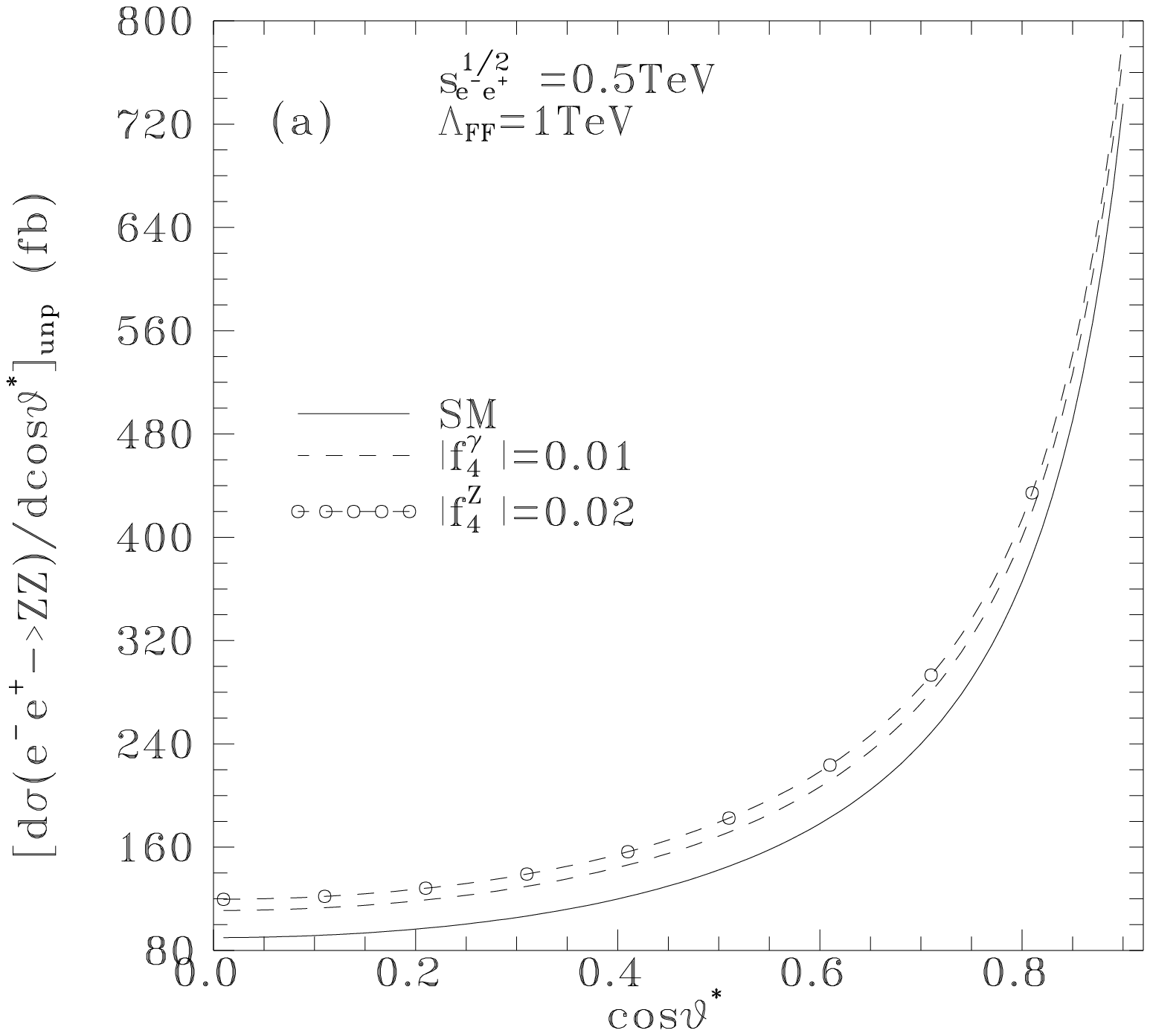,height=7.5cm}\hspace{0.5cm}
\epsfig{file=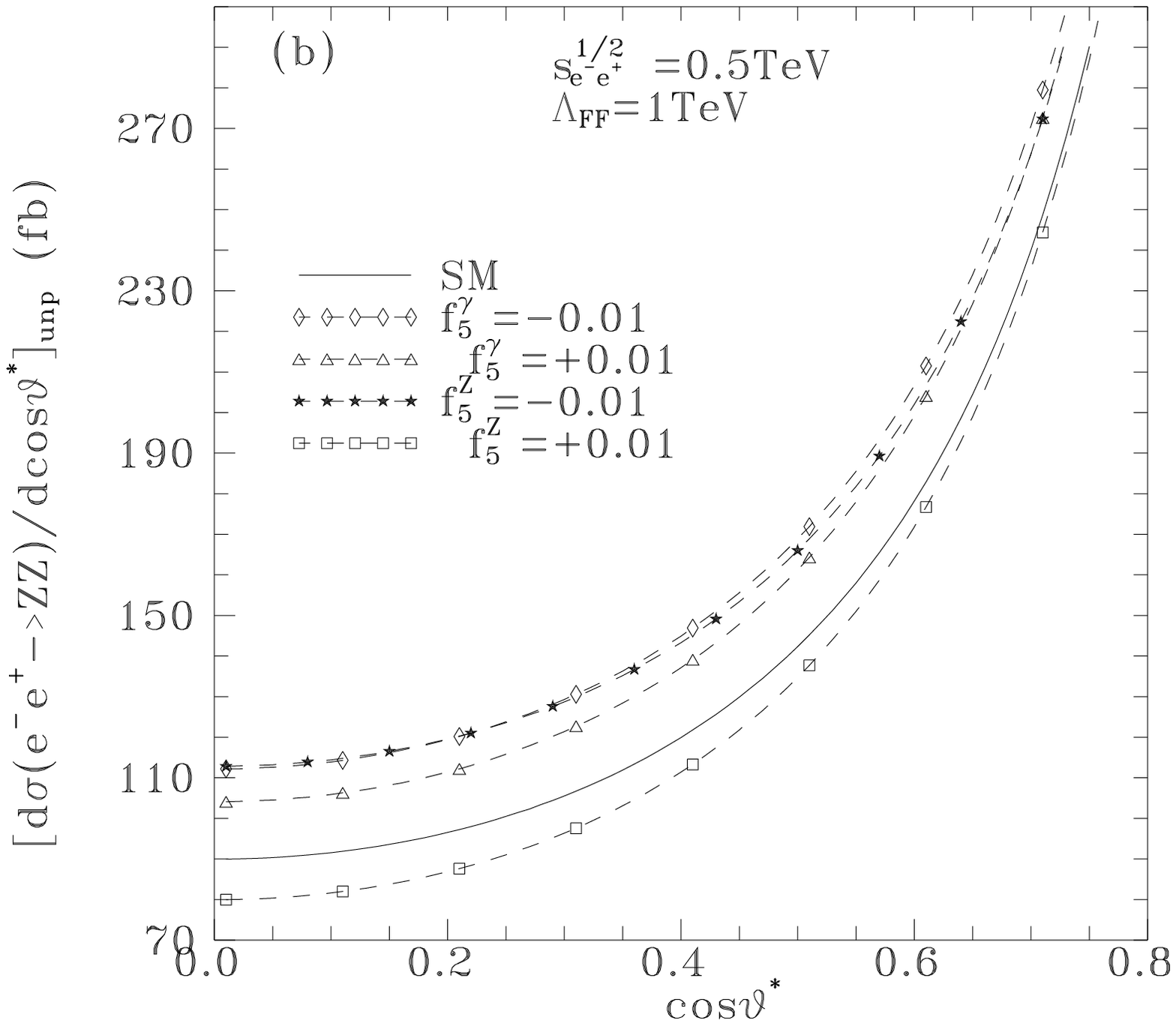,height=7.5cm}
\]
\vspace*{0.5cm}
\caption[1]{Standard and anomalous contributions to unpolarized
$e^-e^+ \to ZZ $ cross sections at an LC.}
\label{LC-fig-ZZ}
\end{figure}

\begin{figure}[p]
\vspace*{-4cm}
\[
\epsfig{file=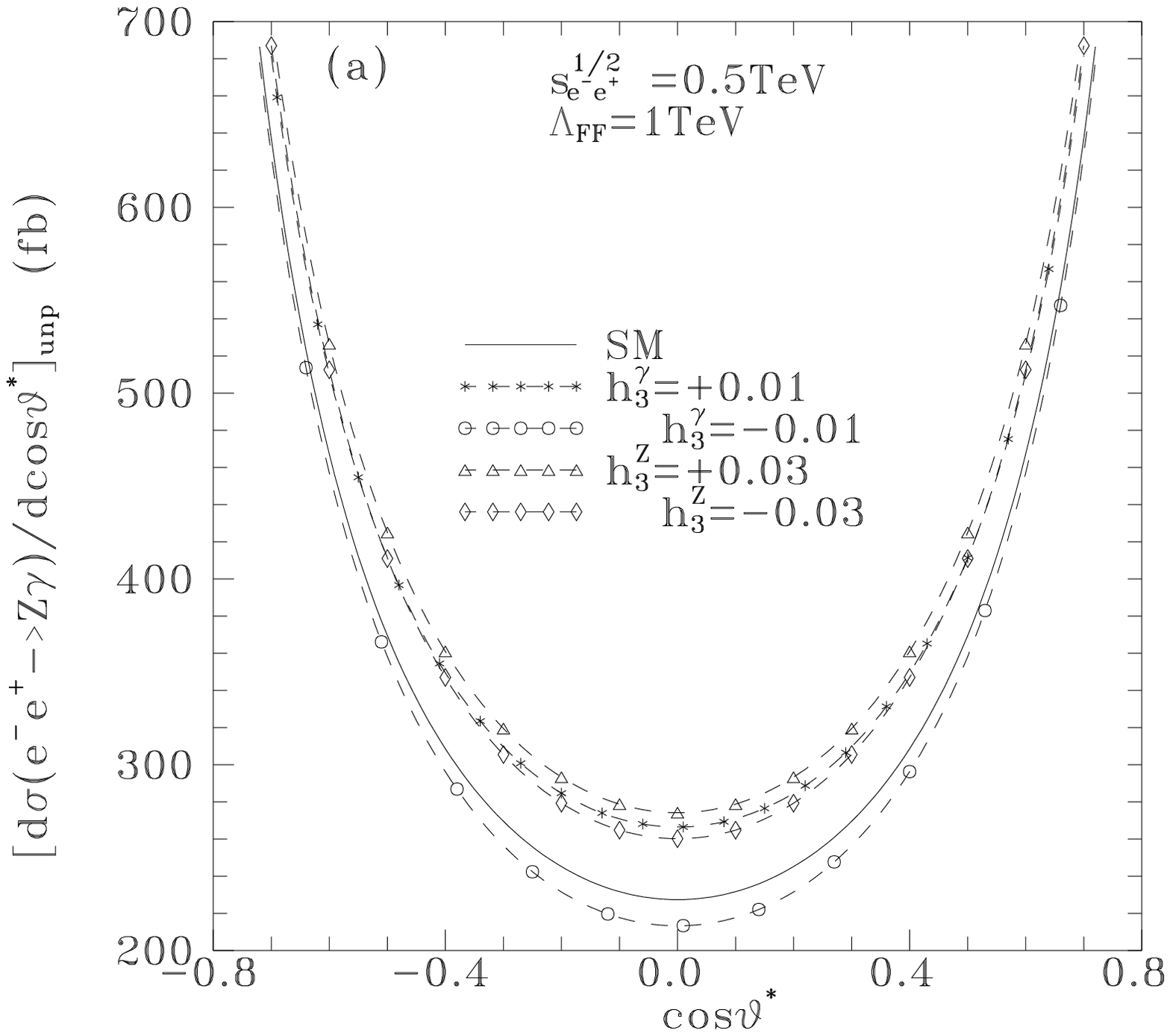,height=7.5cm}\hspace{0.5cm}
\epsfig{file=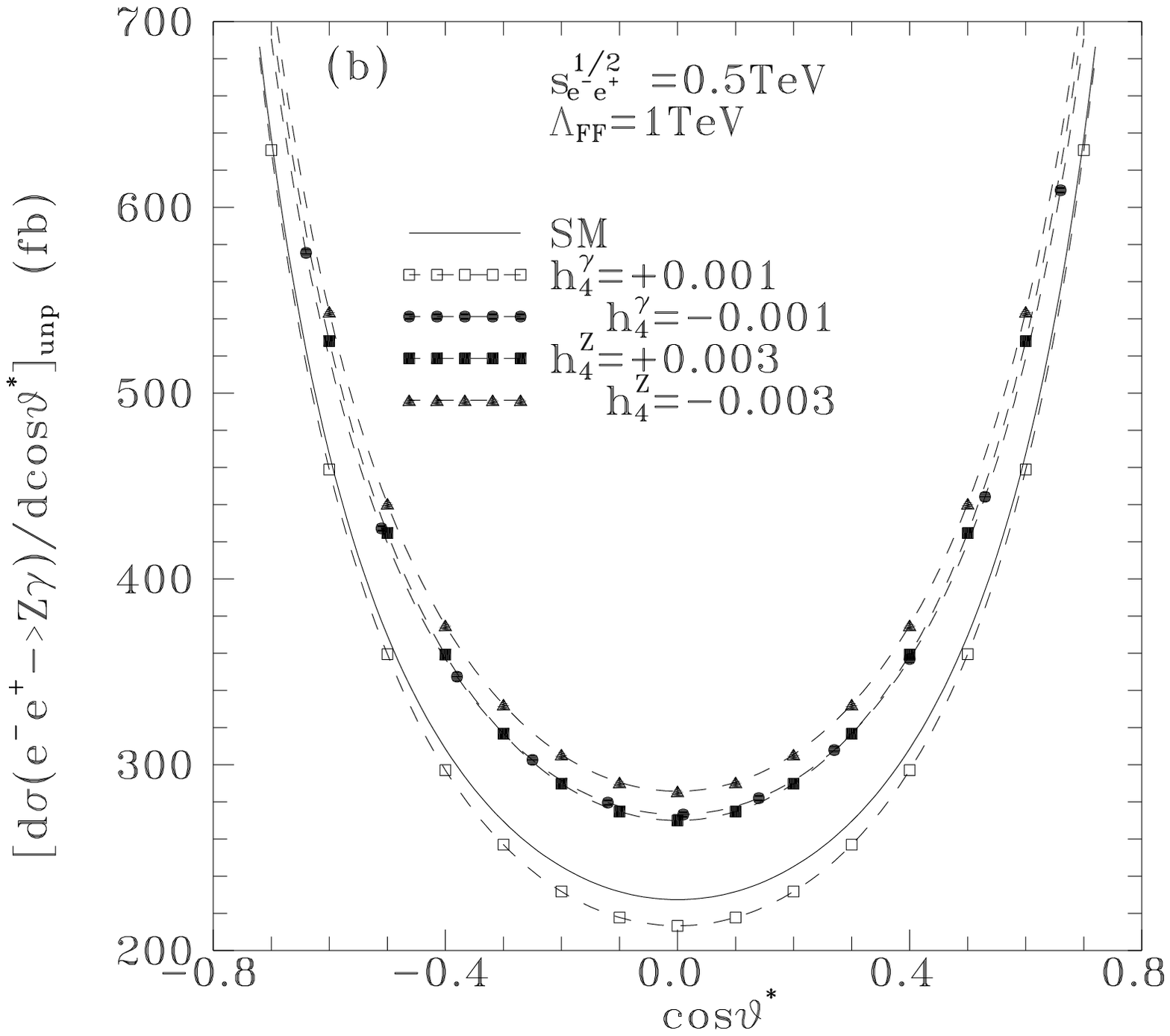,height=7.5cm}
\]
\vspace*{0.5cm}
\[
\epsfig{file=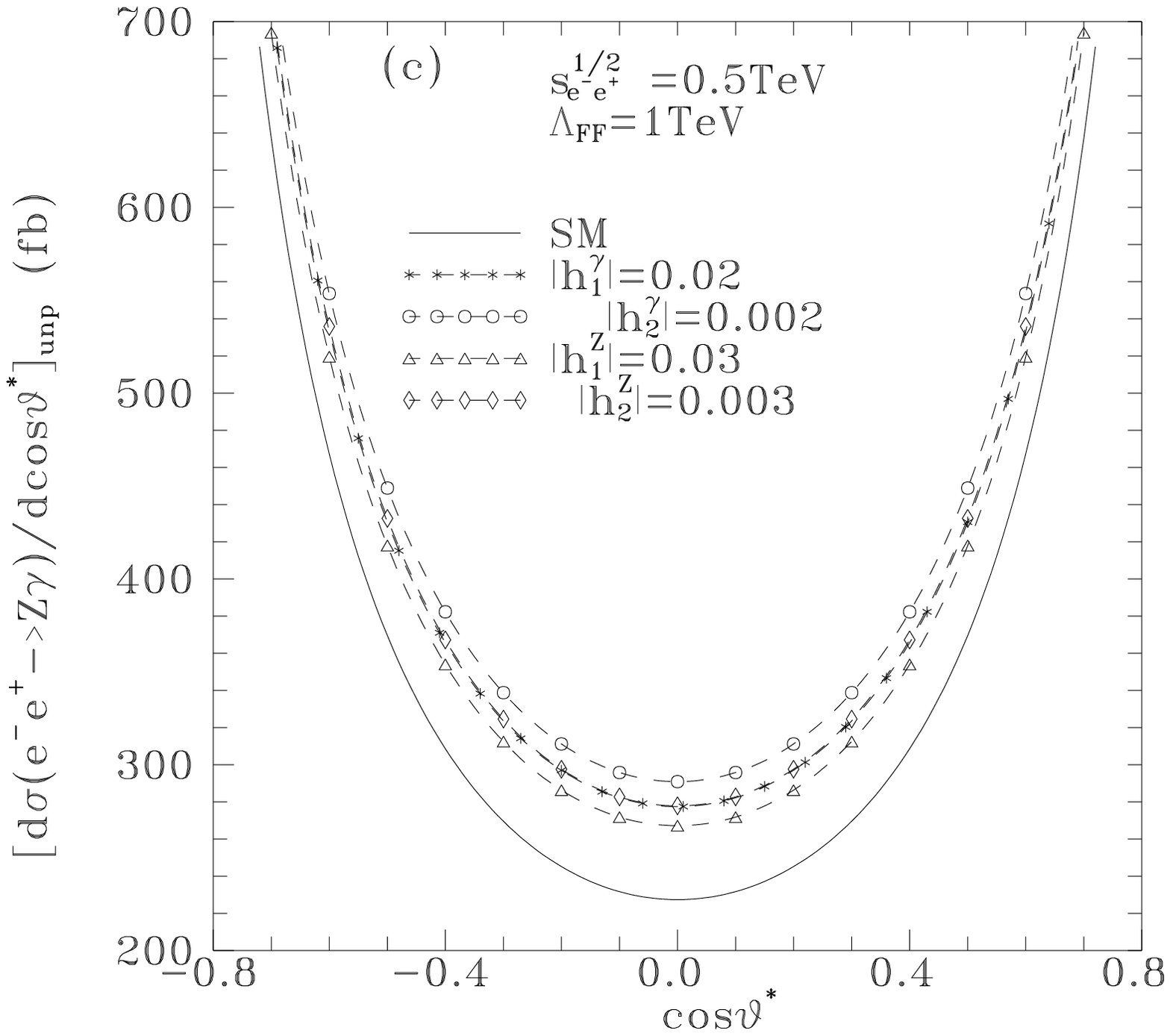,height=7.5cm}
\]
\vspace*{0.5cm}
\caption[1]{Standard and anomalous contributions to unpolarized
$e^-e^+ \to Z\gamma$ cross sections at an LC.}
\label{LC-fig-Zg}
\end{figure}

We present an illustration in
Fig.\ref{LC-fig-ZZ}a,b for $ZZ$ and Fig.\ref{LC-fig-Zg}a,b,c
for $Z\gamma$, by choosing values for
the couplings which make the SM and SM+NP curves
well visible on the drawing; but of course  the observability limits
are found to  correspond to much lower values. To be more precise,
a careful study of the background should be done. One can find
some preliminary studies of these effects presented at the ECFA
meeting \cite{Alcaraz}. In the case of $e^+e^-\to ZZ$ there is
almost no background for the $q\bar q l \bar l$ mode, but there is
some background in the $q\bar q \nu \bar \nu$ mode due to the $WW$
channel. Taking them into account, a final (statistical + systematical)
accuracy of the order of 1\%  should be expected, for a conservative
integrated luminosity of $100~fb^{-1}$. We may even expect a better
sensitivity with the higher luminosity of the TESLA design.
In any case a 1\% accuracy on the cross section at large angles
(see Fig.\ref{LC-fig-ZZ}a,b), would lead to sensitivity limits for
$f^{\gamma}_4$, $f^Z_4$, $f^{\gamma}_5$, $f^{Z}_5$ like
 $2.10^{-3},~4.10^{-3},~3.10^{-3},~7.10^{-4}$ respectively,
at the one standard deviation level.\par

In the case of $e^+e^- \to Z\gamma$ at large angles ($|cos\theta|<0.8$)
no appreciable background is expected \cite{Alcaraz}. With
$100~fb^{-1}$,
about $50000$ events should be selected, leading to
an accuracy better than the 0.5\% level. The sensitivity
(one standard deviation) is now of
$3.10^{-3}$, $5.10^{-3}$, $3.10^{-4}$, $4.10^{-4}$,
$2.10^{-4}$, $4.10^{-3}$, $4.10^{-5}$, $3.10^{-4}$,
for
$h^{\gamma}_1$, $h^Z_1$, $h^{\gamma}_2$, $h^Z_2$,
$h^{\gamma}_3$, $h^Z_3$, $h^{\gamma}_4$, $h^Z_4$, respectively.
Indeed, the observability limits should be
about two orders of magnitude better than the
ones quoted in the LEP2 case.\par
Another feature of LC is the possibility of having longitudinally
polarized $e^{\pm}$ beams; (a polarized $e^-$ beam would be in fact
sufficient, like at SLC).
We have therefore looked at the effect of the anomalous couplings
on the $A_{LR}$ asymmetry whose expression is given in Appendix B.
Note, from the expression of the SM amplitudes given in Appendix A,
that the SM values of $A_{LR}$ are independent of the scattering angle
and energy,   taking the values
\bqa
A_{LR}^{SM}(e^-e^+\to ZZ)& = &
{(g^Z_-)^4-(g^Z_+)^4\over(g^Z_-)^4+(g^Z_+)^4} \simeq 0.28  ~~ ,
\label{ALR-ZZ-SM}\\
A_{LR}^{SM}( e^+e^-\to Z\gamma)
& = & {(g^Z_-)^2 -(g^Z_+)^2\over (g^Z_-)^2 +(g^Z_+)^2}
\simeq 0.14  ~~ . \label{ALR-Zg-SM}
\eqa

\begin{figure}[p]
\vspace*{-4cm}
\[
\epsfig{file=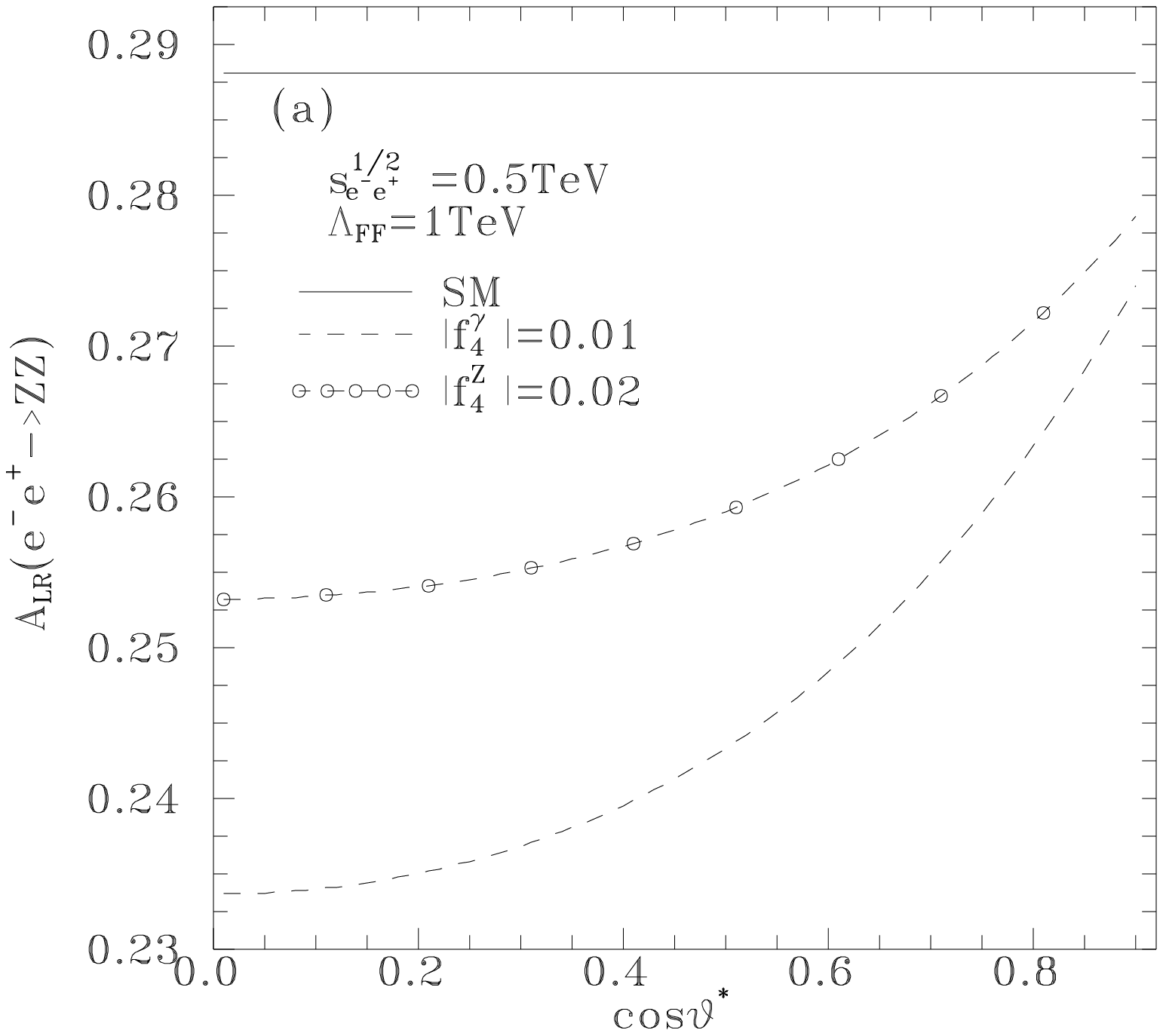,height=7.5cm}\hspace{0.5cm}
\epsfig{file=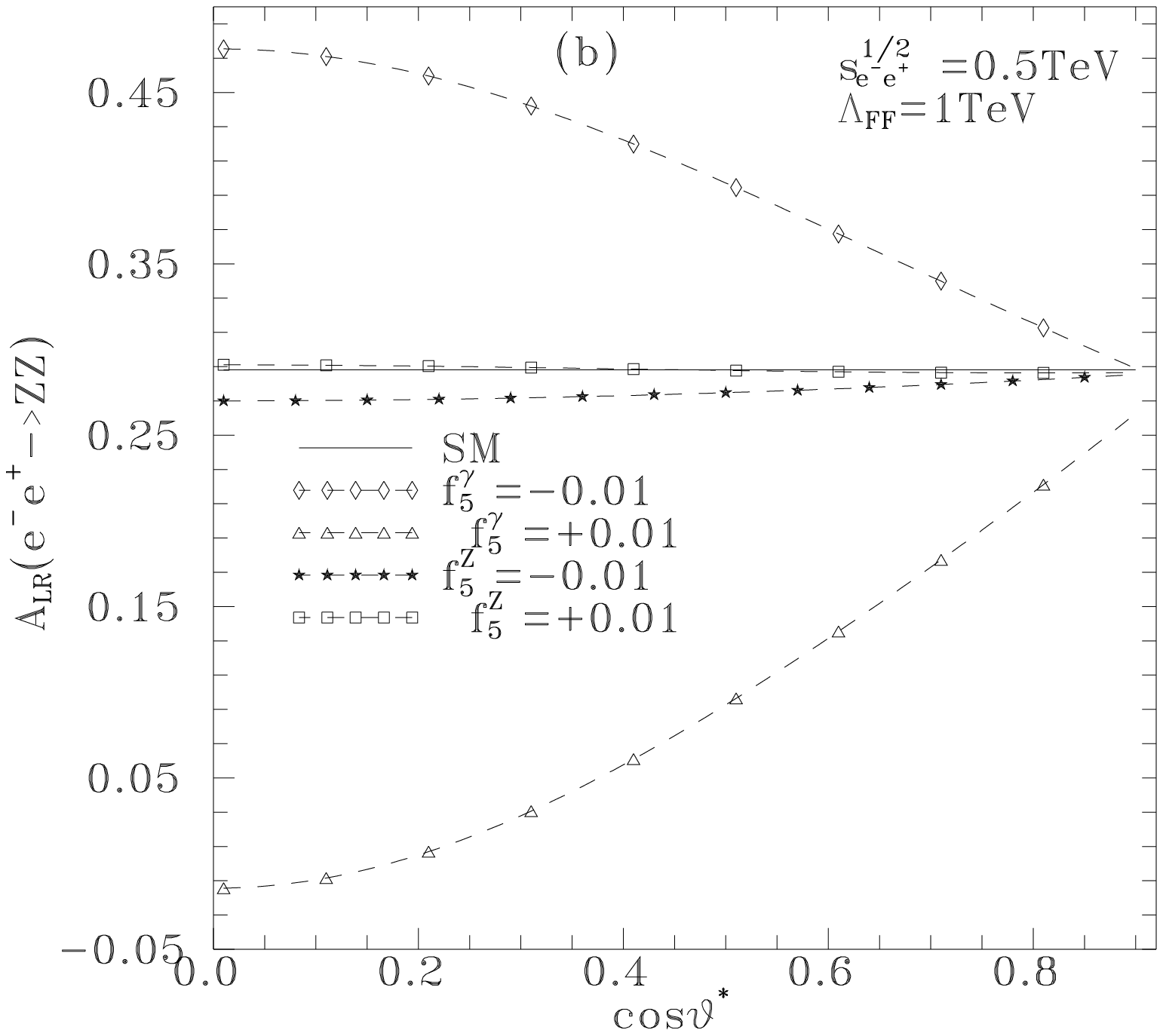,height=7.5cm}
\]
\vspace*{0.5cm}
\caption[1]{Standard and anomalous contributions to
Left-Right asymmetries in the
$e^-e^+ \to ZZ $ cross sections at an LC.}
\label{ALR-fig-ZZ}
\end{figure}

\begin{figure}[p]
\vspace*{-4cm}
\[
\epsfig{file=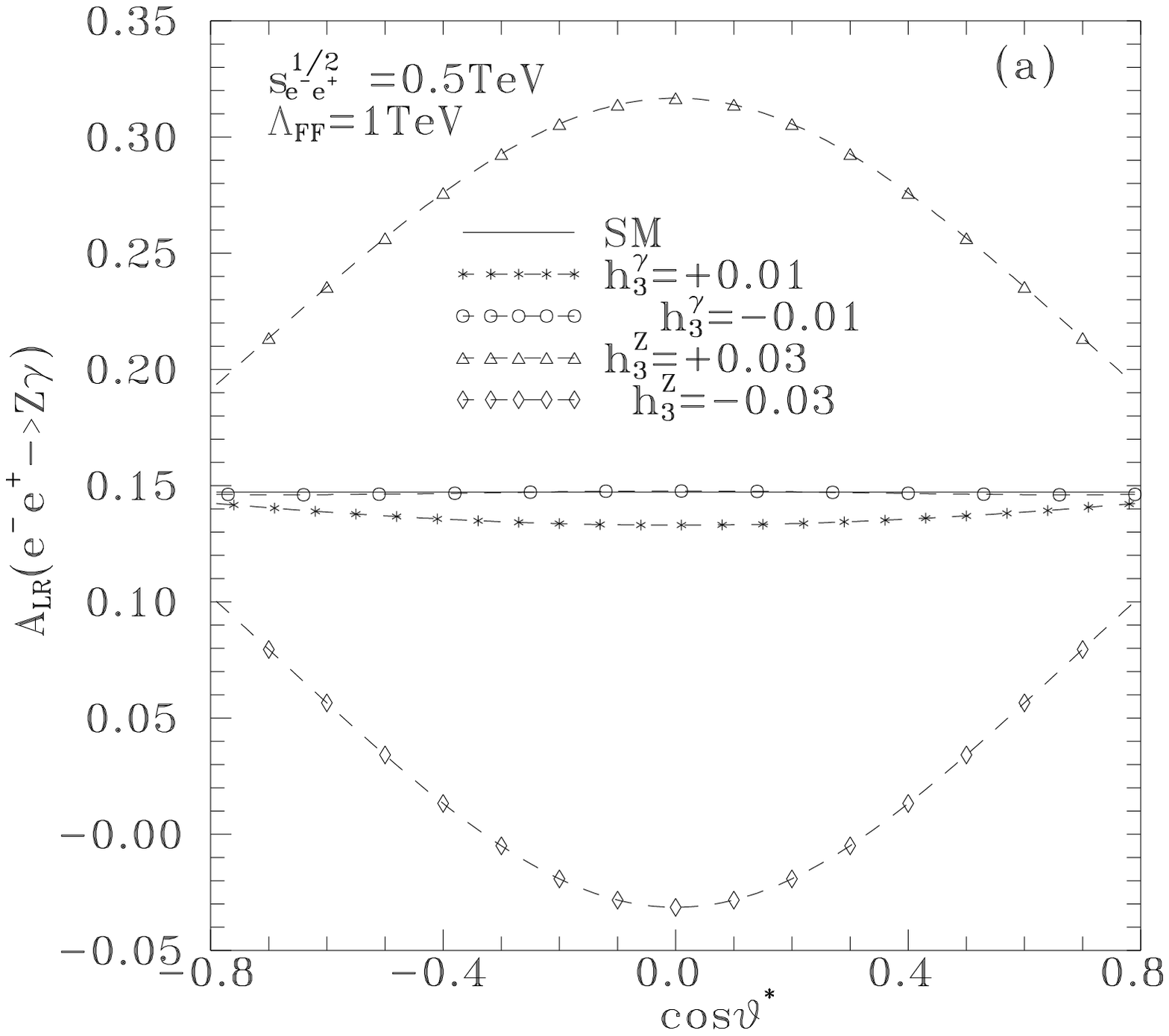,height=7.5cm}\hspace{0.5cm}
\epsfig{file=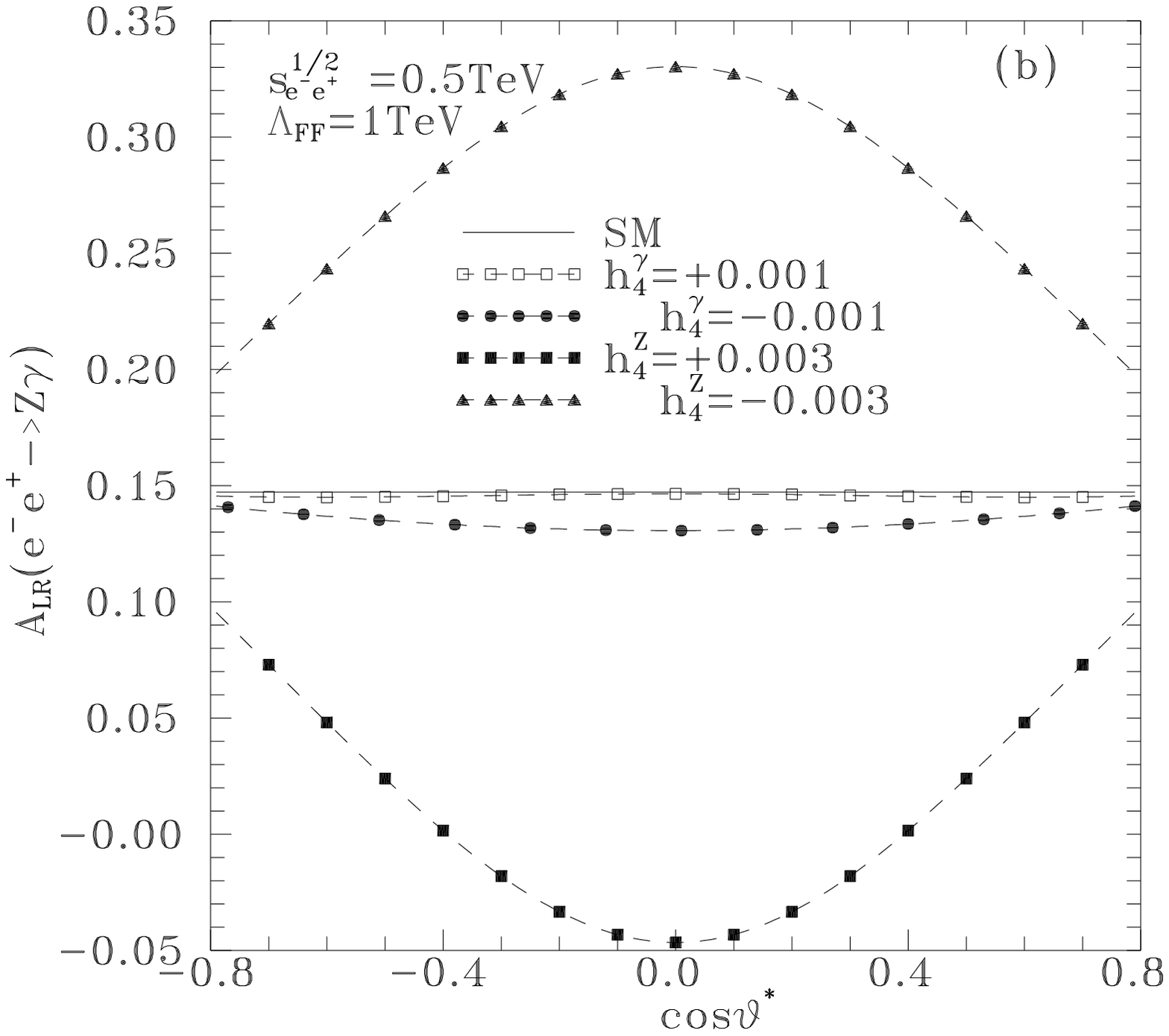,height=7.5cm}
\]
\vspace*{0.5cm}
\[
\epsfig{file=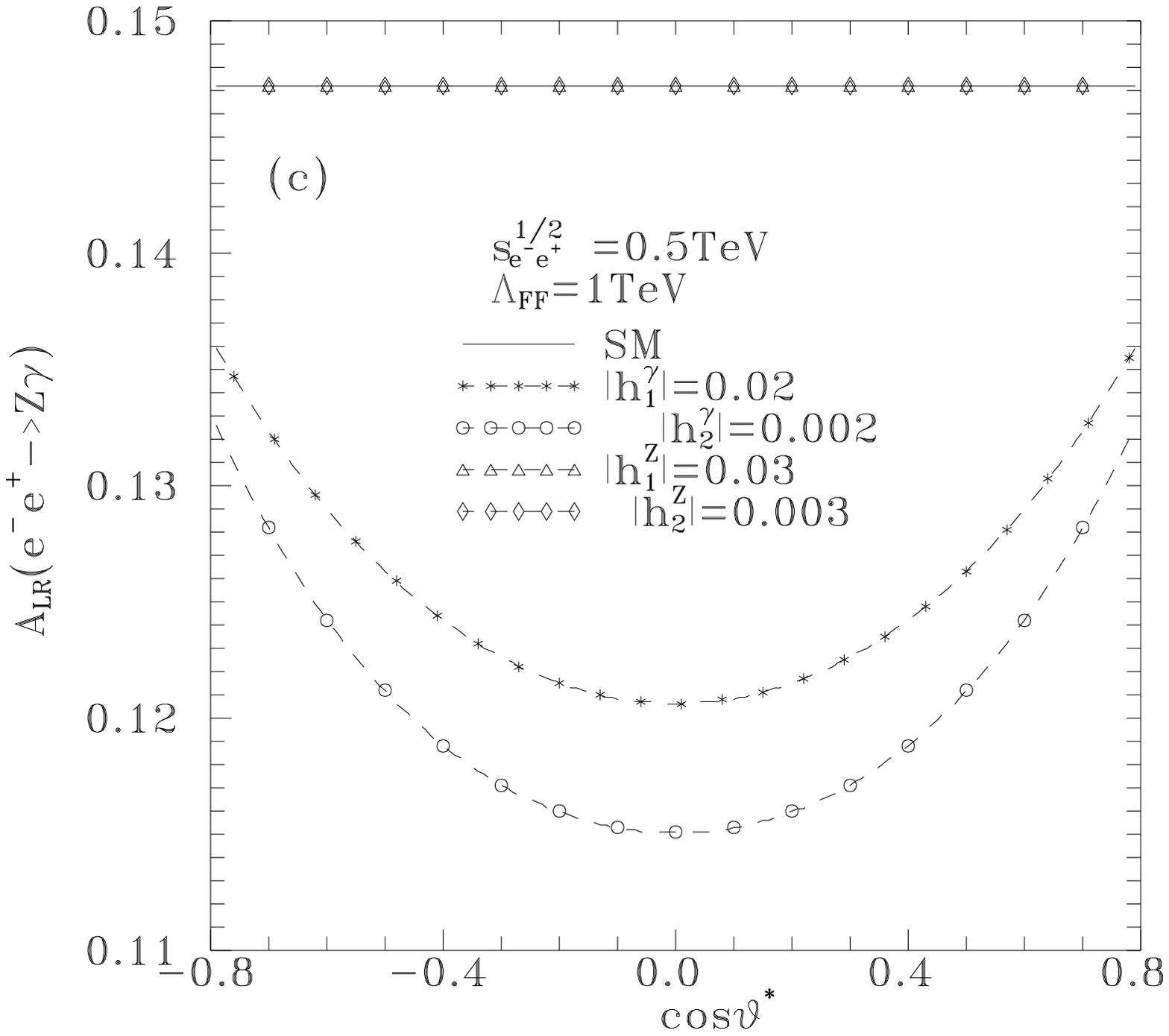,height=7.5cm}
\]
\vspace*{0.5cm}
\caption[1]{Standard and anomalous contributions to
Left-Right asymmetries in the
$e^-e^+ \to Z\gamma$ cross sections at an LC.}
\label{ALR-fig-Zg}
\end{figure}

NP departures from there relations
arise very differently for the anomalous photon- and
$Z$-couplings; especially in the CP-conserving case
which interferes with SM. Thus, one observes a large sensitivity
to the sign of certain  anomalous CP conserving   couplings.
See Fig.\ref{ALR-fig-ZZ}a,b for $ZZ$
and Fig.\ref{ALR-fig-Zg}a,b,c for $Z\gamma$.
It appears therefore, that measurements of $A_{LR}$ should be very
useful for disentangling the anomalous photon and Z couplings.\par

\begin{figure}[p]
\vspace*{-4cm}
\[
\epsfig{file=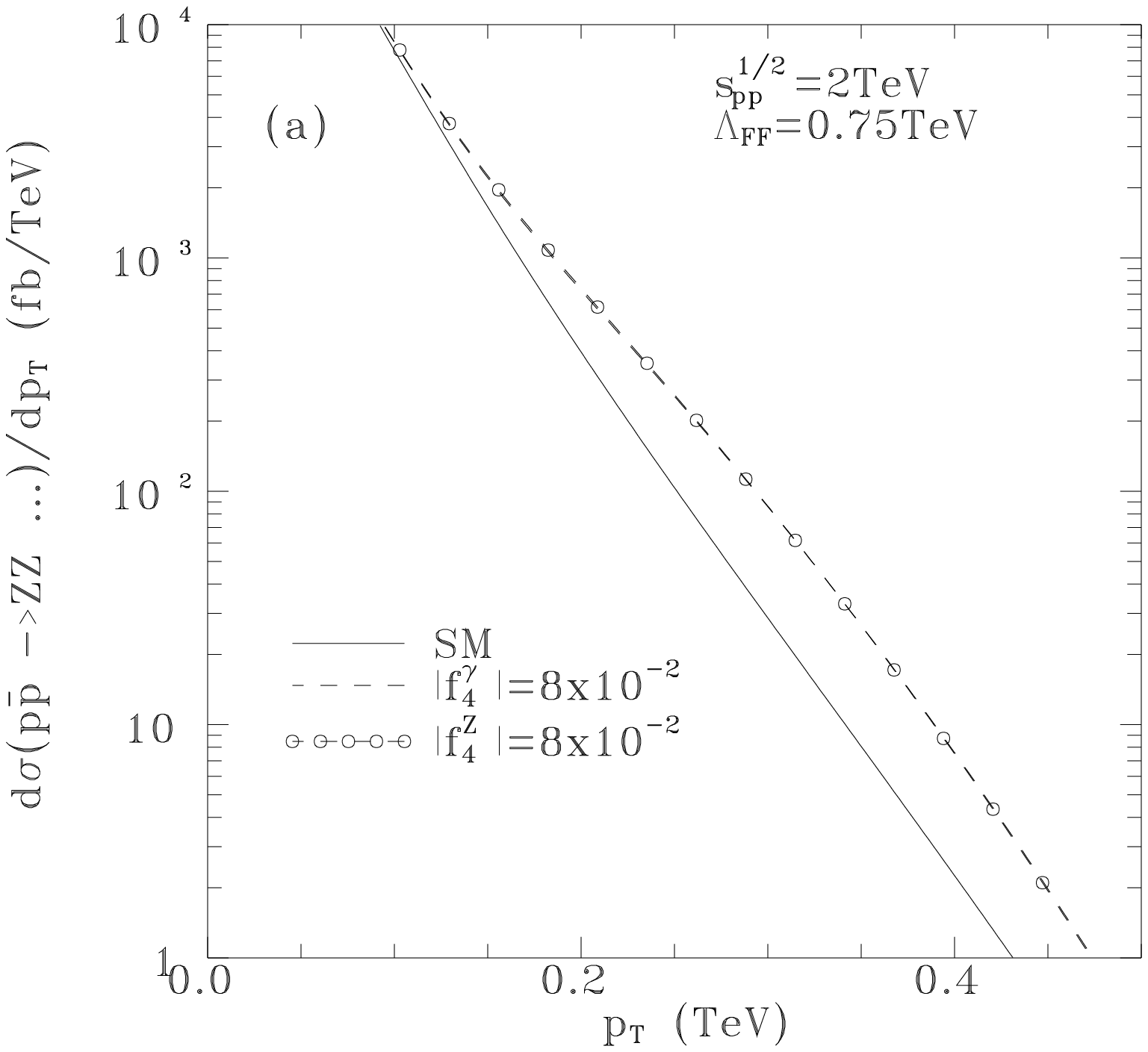,height=7.5cm}\hspace{0.5cm}
\epsfig{file=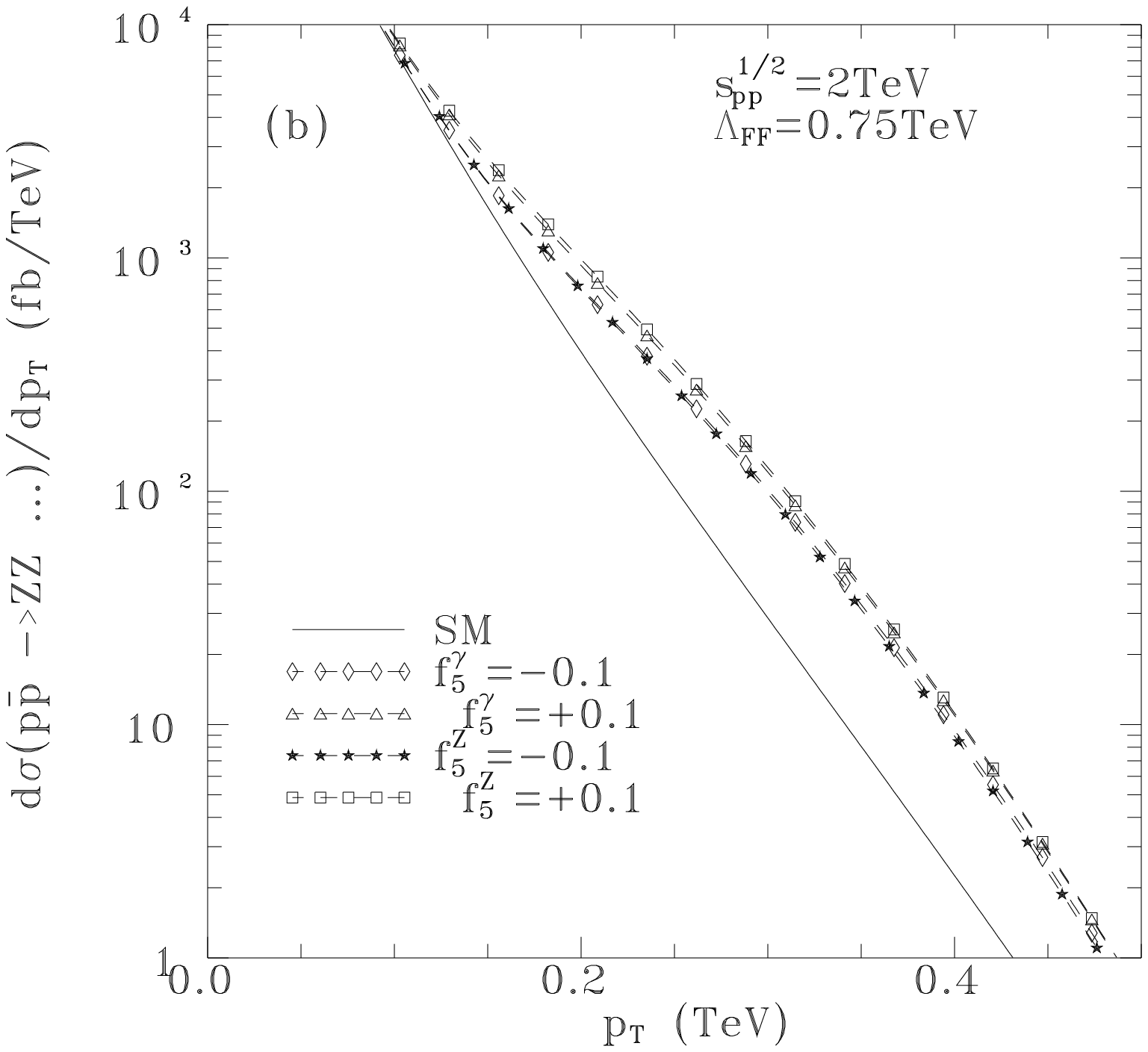,height=7.5cm}
\]
\vspace*{0.5cm}
\[
\epsfig{file=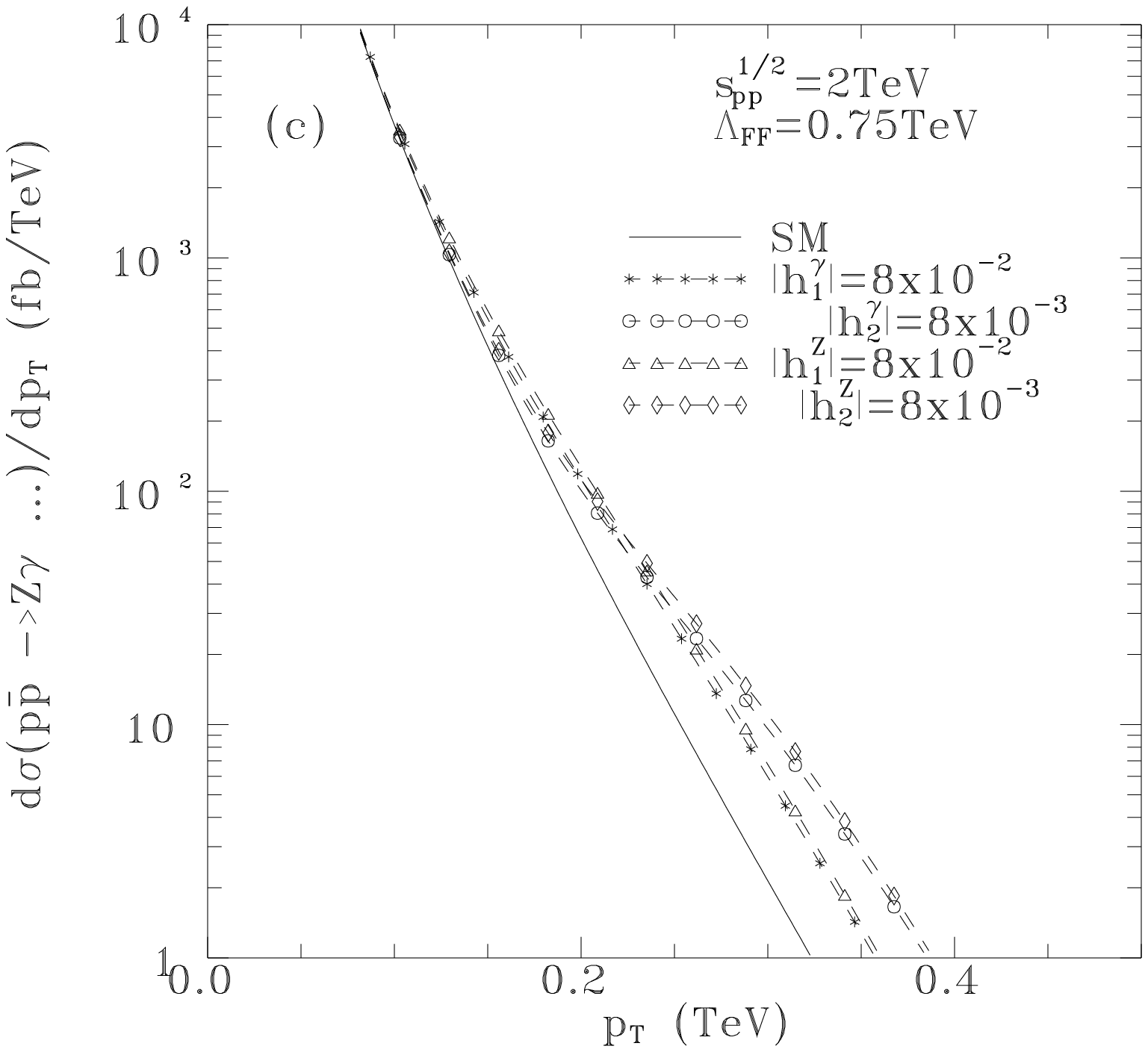,height=7.5cm}\hspace{0.5cm}
\epsfig{file=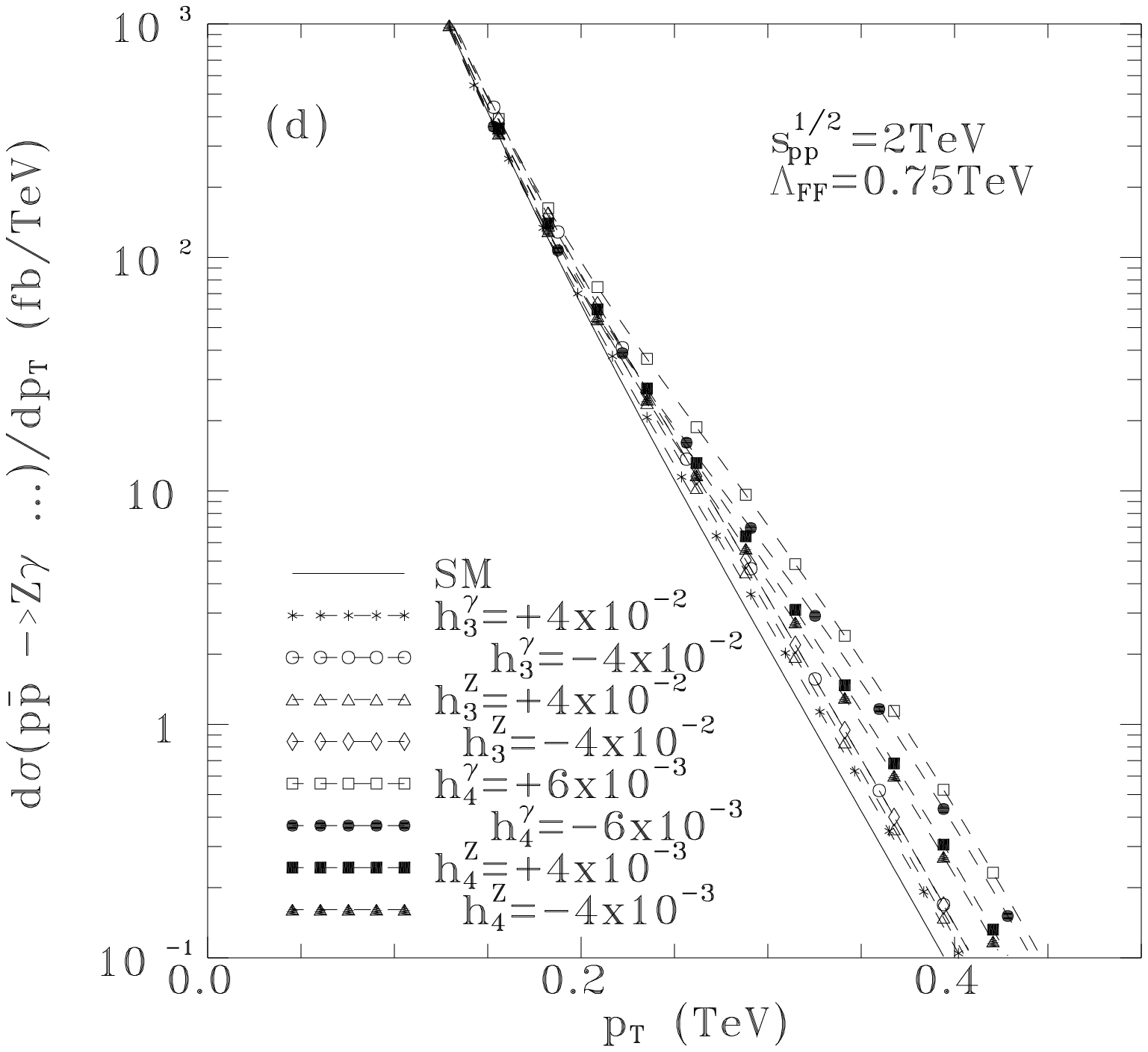,height=7.5cm}
\]
\vspace*{0.5cm}
\caption[1]{Standard and anomalous contributions to
the $p p  \to ZZ ~,~  Z\gamma$ inclusive cross sections at
the Tevatron.} \label{Tev-fig}
\end{figure}

\begin{figure}[p]
\vspace*{-4cm}
\[
\epsfig{file=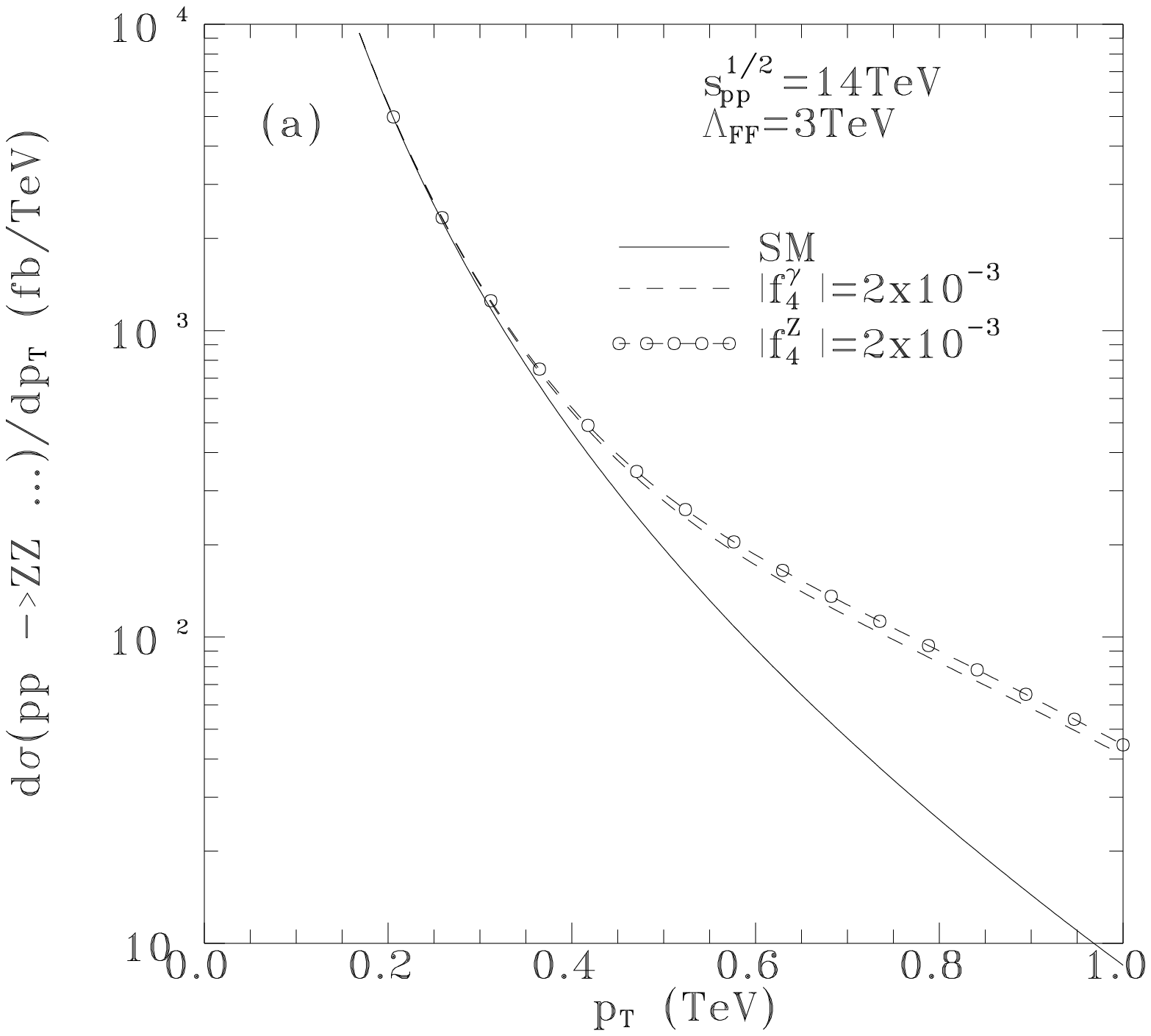,height=7.5cm}\hspace{0.5cm}
\epsfig{file=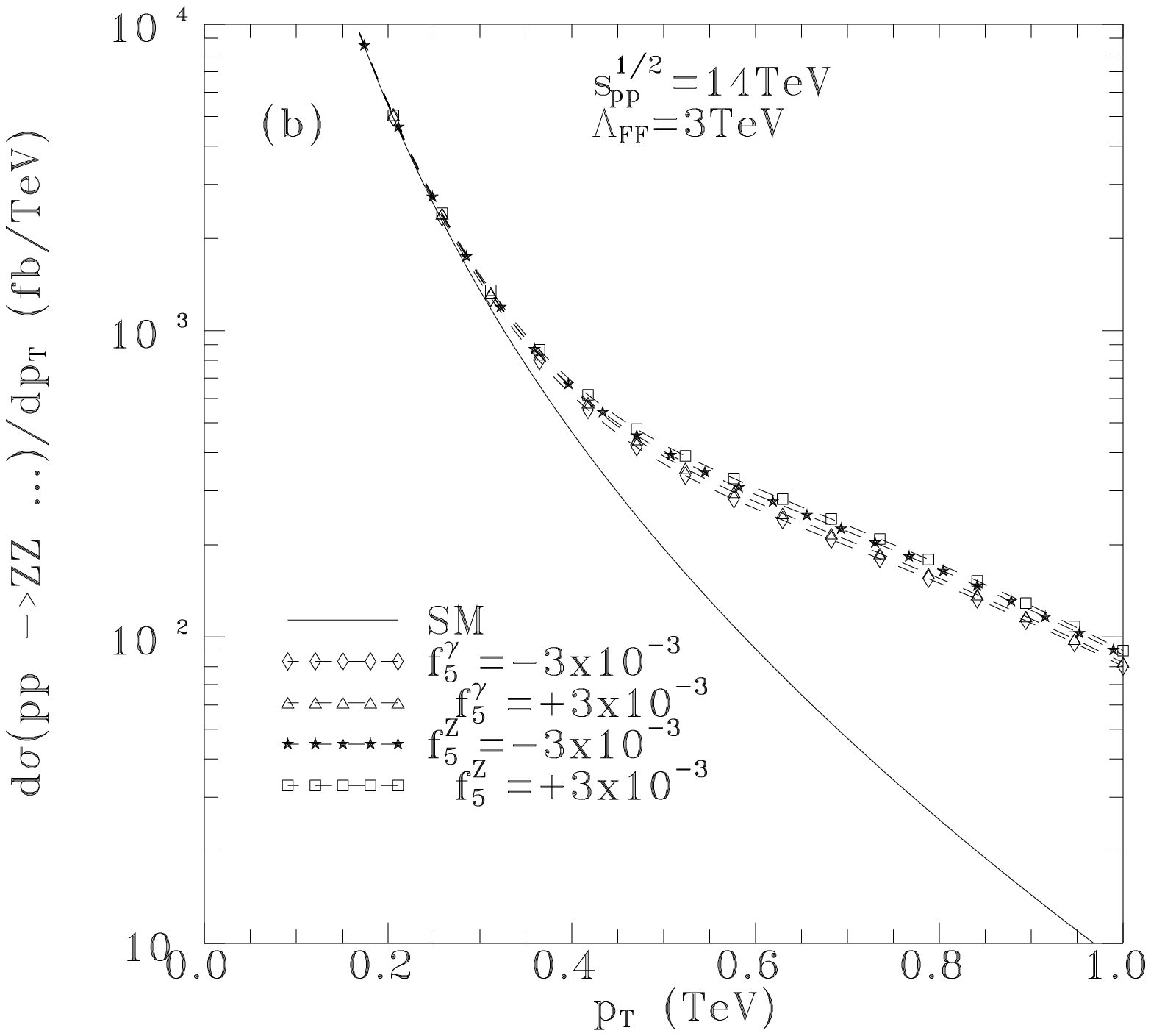,height=7.5cm}
\]
\vspace*{0.5cm}
\[
\epsfig{file=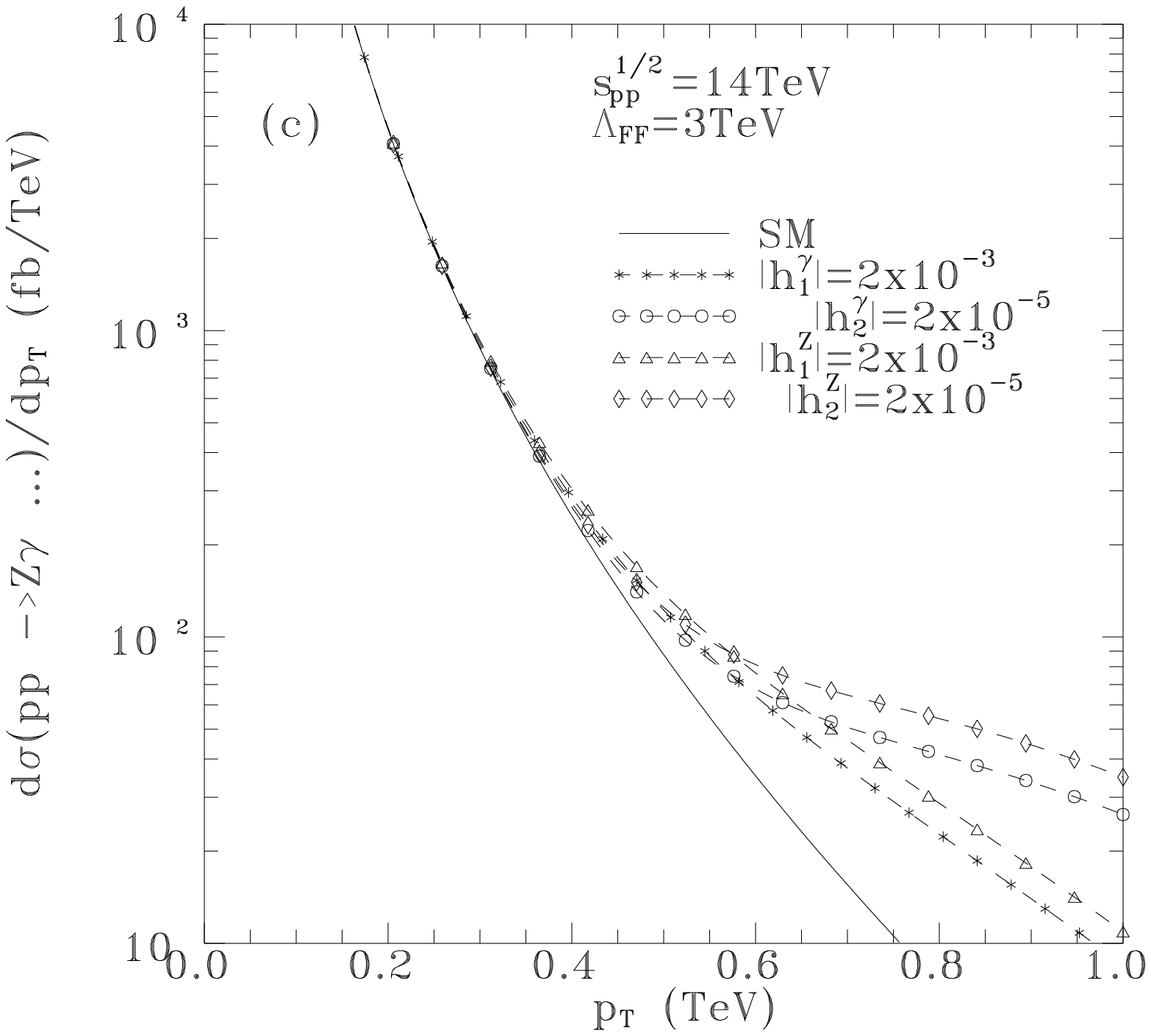,height=7.5cm}\hspace{0.5cm}
\epsfig{file=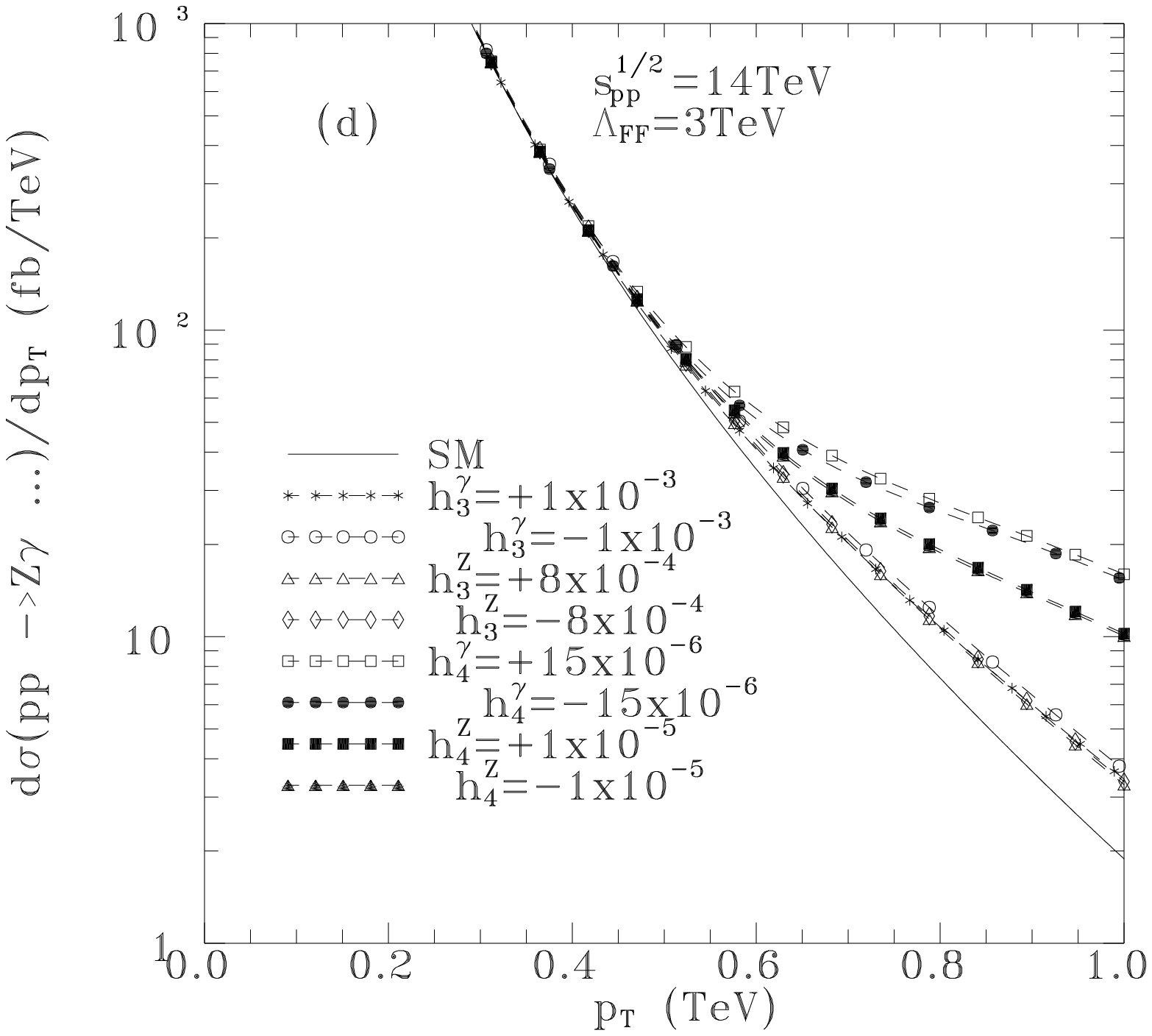,height=7.5cm}
\]
\vspace*{0.5cm}
\caption[1]{Standard and anomalous contributions to
the $p p  \to ZZ  ~,~ Z\gamma$ inclusive cross sections at
LHC.} \label{LHC-fig}
\end{figure}

\vspace{0.5cm}
\noindent
{\bf Application to $ZZ$ and $Z\gamma$ production at hadron
colliders.}\\
Finally we have made an illustration for $ZZ$ and $Z\gamma$ at the
TEVATRON (2TeV) and at LHC (14TeV).
We have chosen to illustrate the transverse momentum ($p_T$)
distribution of one Z (both in the $ZZ$
and in the $Z\gamma$ case), as it is the one which is mostly used
in the literature \cite{Baur}.
But we have also checked that the $ZZ$ or $Z\gamma$ invariant mass
($\sqrt{\hat s}$)
distribution shows roughly the same features and gives the same
sensitivity to the anomalous couplings.
These distributions reflect the fact that CP-conserving
 amplitudes interfere
with the SM ones, whereas the CP-violating ones always do not,
as we can see in Fig.\ref{Tev-fig} for the TEVATRON  and
Fig.\ref{LHC-fig} for the LHC .\par

These interference
patterns are somewhat less pronounced than in the
illustrations for  $e^-e^+$ collisions. This is
due to the fact that the $p_T$ distributions that we are showing,
are the results of integrations over regions of
phase space  where the quadratic term is important
and  partly washes out the
interference term. In order to recover the same features as in
the $e^-e^+$ illustrations, one should make severe cuts selecting a
restricted domain in invariant mass and preferably
large values of the cm scattering angle. In such a domain one can find
values of the CP-conserving couplings producing a visible effect dominated
by the linear (interfering) term.
Such a study, which should be carefully done taking into account all
the event selection criteria, is beyond the scope of this paper, but
we think that it should be tried. For this purpose we have given in
Appendix B the expression of the differential cross section with
respect to invariant mass and c.m. scattering angle.\par

As far as the comparison of the sensitivities at hadron colliders and
at $e^-e^+$ colliders is concerned, we would like to come back to
the discussion we gave at the begining of this Section about the use of
form factors, by adding a few
(more or less obvious) remarks. This  use of
form factors is commonly done
when analyzing transverse momentum or invariant mass
distributions at hadron colliders, in order to take into account
the fact that any given non vanishing value of an anomalous
coupling, will eventually violate unitarity at sufficiently
higher energies. In spite of its apparent necessity, this procedure
forbids to do any clear comparison of observability limits among
different colliders because they involve an integration over a large
range of invariant mass $\sqrt{\hat s}$.
We would therefore prefer another procedure that
would consist in giving observability limits for the
considered basic couplings (without any form factor),
in restricted domains (bins) of the subprocess invariant mass.
At the same collider, one could then  establish a set
of different observability limits, by taking a set of such domains
of invariant masses in which there are enough events to analyze.
This set of observability limits could then be compared
among each other and also with observability limits obtained
separately at different colliders and different energies.\par

This is an additional motivation for our suggestion to make
analyses in restricted domains of invariant masses, that we already
mentioned above.

\section{Final discussion}

In this paper we have examined the existing phenomenological
description of the anomalous neutral 3-boson couplings ($\gamma\gamma
Z,~\gamma ZZ,~ ZZZ$), and we have reviewed the basic assumptions
which allow to constrain the number and  structure of the relavant
couplings.\par

A first observation was that in the set of couplings which were
commonly used for studying $Z\gamma$ production,
a factor "i" was missing
making the Effective Lagrangian antihermitian. As a result, the
interference patterns of the CP-conserving and the CP-violating NP
amplitudes, with the SM ones, were reversed. Nevertheless, this
observation does not seem to invalidate most of the presently
existing experimental observability limits, since they are so low,
that they are  mainly arising from the quadratic part of the NP
contribution. But of course, as the accuracy of
the measurements is increasing, it would eventually lead to
non-intuitive results.\par

We have therefore carefully rederived the Standard
Model (SM) and the anomalous (NP) amplitudes,
 in order to clearly fix  all conventions and
normalizations. We have illustrated the corresponding
effects that the various anomalous couplings induce on $ZZ$ and
$Z\gamma$ production at $e^-e^+$ and  hadron colliders.
We have made applications for LEP2, for an LC of 500 GeV, and
 for the TEVATRON and the LHC. On the angular distribution,
we have shown the interference patterns
produced by the CP-conserving couplings and the additive
contributions given by the CP-violating couplings.\par

 In the case
of LC, we have emphasized the possible role of longitudinal
polarization for disentangling different sets of couplings, as in some
cases the interference pattern in the $A_{LR}$ asymmetry
is much more pronounced than on the unpolarized cross section.\par

As far as hadron colliders are concerned, we have suggested
to try to make analyses in different restricted domains (bins)
of invariant $ZZ$ or $Z\gamma$ masses, with large values of
the c.m. scattering angle. In such a case, the inteference patterns
should be comparable to the ones observable in $e^-e^+$ collisions,
and much more pronounced than in a (fully integrated) transverse
momentum distribution. Of course such an analysis
was not possible in the $1.8~TeV$ TEVATRON; but it may be possible,
due to the much larger expected statistics, at the upgraded
TEVATRON  and the LHC.\par

This procedure would also allow
to get rid of the multiplicative form factor introduced in many
previous analyses. The comparison of the various observability
limits obtained at different $e^-e^+$ or hadron colliders, each one
being defined for a given invariant mass range, would then be
straightforward.\par

We have also mentioned that if we use \eg\@ the linear scalar sector
representation appropriate for a relatively light higgs particle,
then at the level of all possible
$SU(2)\times U(1)$ gauge invariant $dim=6$ operators, which
predict a certain pattern of anomalous $\gamma WW$ and $ZWW$
couplings, no neutral gauge boson couplings are expected.
Thus if such couplings are discovered, it may mean either
that for some reasons certain $dim=8$ or 10 operators are more
important than those of $dim=6$;  or that NP scale is nearby,
thus invalidating the dimensional ordering of the
$SU(2)\times U(1)$ gauge invariant operators.\par

We have made a specific application taking as NP effect the one due
to the contributions of heavy fermions at one loop. We have discussed
two cases, one in which only one set of heavy fermions is lighter than
all others, and one in which the complete family is nearly degenerate.
In both cases one observes that only the $f^V_5$ and $h^V_3$ couplings
are generated (together with the anapole $ZWW$, $\gamma WW$ couplings
in the charged
sector), the other couplings requiring higher order effects.
In addition, we have noticed remarkable relations among these couplings.
An important difference between these two aforementioned cases
is that, in the first one,
the couplings behave like $1/ \Lambda^2_{NP}$, whereas in the
second  they go like  $1/ \Lambda^4_{NP}$, leading to much
poorer bounds on the NP scale. We should also state that,
within this type
of models, in order to generate observable couplings at present
colliders,
the NP dynamics must include a
strong enhancement effect that would compensate
the one loop $\alpha/4\pi$ factor. Otherwise one could expect
such virtual effects to be observable only at a very high
luminosity LC. In any case this example has shown
how experimental constraints established
on each coupling could give some indications on the NP
properties.\par

{\bf Acknowledgments}:
It is a pleasure to thank Robert Sekulin
for very informative discussions and suggestions.\\

\newpage

\renewcommand{\theequation}{A.\arabic{equation}}
\renewcommand{\thesection}{A.\arabic{section}}
\setcounter{equation}{0}
\setcounter{section}{0}

{\large \bf Appendix A: The Standard Model helicity amplitudes
 for $f \bar f \to Z Z $ and  $f \bar f \to Z \gamma $.}\\

The invariant helicity amplitudes for the production processes of the
neutral vector bosons $V_1,V_2$,
\bq
f (k_1, \lam_1) \bar f(k_2, \lam_2)
\to V_1(q_1, \tau_1) V_2(q_2, \tau_2) ~~ ,
\label{ffV1V2}
\eq
are denoted  as\footnote{Its sign is related to the sign of
the S-matrix through
$S_{\lam_1 \lam_2 \tau_1 \tau_2}= 1 + i (2\pi)^4
\delta(p_f-p_i) F_{\lam_1 \lam_2 \tau_1 \tau_2}$ .}
$ F_{\lam_1 \lam_2 \tau_1 \tau_2}$, where
 the momenta and helicities of the incoming fermions
($f, \bar f$) and the outgoing neutral vector bosons
are indicated in parentheses in (\ref{ffV1V2}).
Since at Collider energies
the mass $\mf$ of the incoming fermion can be neglected in all
cases\footnote{Except for the top quark of course, which is of no
  relevance here.},   the dependence of the amplitude on the
initial helicities is only through the combination
$\lam \equiv \lam_1- \lam_2$, so that the  notation
 $F^\lam_{\tau_1 \tau_2}$  will be used below. Thus,
there two possible values for
 $\lam$ are $\lam = -1 (+1)$, corresponding to
an $f_{L(R)}$ fermion  interacting with a $\bar f_{R(L)}$
antifermion respectively; so that the $Z$ and photon couplings
in (\ref{gphoton},\ref{gZ}) may be written as
$g^Z_\lam$  and $g^\gamma_\lam$  respectively, as defined by the
Standard Model interaction Lagrangian involving  a fermion  $f$
of charge $Q_f$ and third isospin component  $t_f^{(3)}$,

\bq
\L = -e  V^\mu \bar f
\left  (g^V_L \gamma_\mu \frac{(1-\gamma_5)}{2}  ~+ ~
g^V_R \gamma_\mu \frac{(1+\gamma_5)}{2} \right ) f  ~~ ,
\label{Vff}
\eq
with
\bq
g^\gamma_L= g^\gamma _R =Q_f  ~~~,  \label{gphoton}
\eq
\bq
g^Z_L =\frac{1}{\sw\cw}(t_f^{(3)}- Q_f \swd ) ~~~ ,~~~
g^Z_R =\frac{1}{\sw\cw}(- Q_f \swd ) \label{gZ}
\eq

For $ ZZ$  production, CPT invariance implies at tree level
\bq
F^\lam_{\tau_1, \tau_2}( f \bar f \to  ZZ)=
F^{\lam *}_{-\tau_2 ,-\tau_1}( f \bar f \to  ZZ) ~~, \label{CPT-ZZ}
\eq
\noindent
while CP invariance would demand
\bq
F^\lam_{\tau_1, \tau_2}( f \bar f \to  ZZ)=
F^{\lam }_{-\tau_2 ,-\tau_1}( f \bar f \to  ZZ) ~~ , \label{CP-ZZ}
\eq
even at higher orders. Since the  Standard Model
amplitudes for $f\bar f \to ZZ$ satisfy CP invariance,
the tree level standard helicity amplitudes are real and
may be written, following the notations of \cite{Hagiwara}, as
\bq
F^\lambda_{\tau_1 \tau_2}( f \bar f \to  ZZ~;~SM)
= - e^2(g^Z_\lam)^2 ~
\frac{A^{\lam Z}_{\tau_1 \tau_2 }}
{4\beta^2\sin^2\vartheta^* +\gamma^{-4}}
 ~~ , \label{FffZZ}
\eq
where
\bqa
&& A^{\lambda Z}_{\tau_1, \tau_2} =
2\sin\vartheta^*
\left [ 2 \lam \cos\vartheta^* (\beta^2-\tau_1 \tau_2)
-(1+\beta^2)(\tau_2-\tau_1) \right ] ~ \mbox{for} ~
 \tau_1\tau_2 \neq 0 ~, \label{AZtau1tau2} \\
&& A^{\lam Z}_{0,\tau}=A^{\lam Z}_{-\tau, 0}=
\frac{2\sqrt{2}}{\gamma}(1-\lam \tau \cos\vartheta^*)
[\lam \tau (1+\beta^2) +2 \cos\vartheta^*] ~ \mbox{for} ~ \tau\neq 0
~,
\label{AZ0tau} \\
&& A^{\lam Z}_{00} = -~ \frac{4 \lam \sin(2\vartheta^*)}{\gamma^2} ~~ ,
\label{AZ00}
\eqa
and $\vartheta^*$ is the c.m. $Z$ scattering angle with respect to the
$f$-beam axis, while
\bq
\beta =\sqrt{ 1-~\frac{4\mzd}{\s}} ~~~ , ~~~
\gamma=\frac{1}{\sqrt{1-\beta^2}} \label{ZZ-beta-gamma} ~ .
\eq
Notice that (\ref{FffZZ}-\ref{AZ00}) define also our conventions on the
relative  fermion and antifermion  phases.

Correspondingly  the SM helicity amplitudes for $f \bar f \to Z \gamma$
are
\bq
F^\lambda_{\tau_1 \tau_2}( f \bar f \to  Z \gamma ~;~ SM)
= - e^2 g^Z_\lam  g^\gamma_\lam ~
\frac{\s}{\s-\mzd} ~ A^{\lam \gamma}_{\tau_1 \tau_2 } ~~ , \label{FffZg}
\eq
where
\bqa
 A^{\lambda \gamma}_{\tau_1, \tau_2} & = & -~\frac{1}{\sin\vartheta^*}
\Big \{ \tau_2-\tau_1 +\lam \cos\vartheta^* (-1+\tau_1\tau_2)
\nonumber \\
&& +~\frac{\mzd}{\s} [\tau_1+ \tau_2 +
\lam (1+\tau_1 \tau_2) \cos\vartheta^*] \Big \}
 ~, \label{Agtau1tau2} \\
 A^{\lam \gamma}_{0,\tau_2} & = &
2 \sqrt{2} \lam\tau_2 ~\frac{\mz}{\sqrt{\s}}
~.  \label{Ag0tau}
\eqa  \par

\vspace{2cm}


\renewcommand{\theequation}{B.\arabic{equation}}
\renewcommand{\thesection}{B.\arabic{section}}
\setcounter{equation}{0}
\setcounter{section}{0}

{\large \bf Appendix B: Cross section for LEP, LC and the Hadron
Colliders.}\\

The full helicity amplitudes for  $V_1V_2$
($V_1=Z,~V_2=Z,\gamma $) production
through the process (\ref{ffV1V2})
are obtained by adding the SM contributions in
(\ref{FffZZ}, \ref{FffZg}) and the NP ones appearing in
(\ref{FffZZ-NP}, \ref{FffZg-NP1}, \ref{FffZg-NP2}) as
\bq
F^\lambda_{\tau_1 \tau_2}( f \bar f \to  V_1V_2)
=F^\lambda_{\tau_1 \tau_2}( f \bar f \to  V_1V_2 ~;~SM)+
F^\lambda_{\tau_1 \tau_2}( f \bar f \to  V_1V_2 ~;~NP)
~ ,
\eq
and they are normalized  so that the unpolarized  differential
cross sections are given by
\bqa
 \frac{d\hat \sigma(f \bar f \to ZZ) }{d \cos \vartheta^*}
&=&\frac{\s}{2}\left (1-~\frac{4\mzd}{\s}\right )^{{1\over2}}~
\frac{d\hat \sigma(f \bar f \to ZZ) }{d \t}
\nonumber \\
&&
= \frac{1}{128\pi \s}\left (1-~\frac{4\mzd}{\s}\right )^{{1\over2}}
\sum_{\tau_1 \tau_2} \left [ |F^{\lam=1}_{\tau_1\tau_2}|^2
+|F^{\lam=-1}_{\tau_1\tau_2}|^2 \right ] ~, \label{dsig-ffZZ}
\eqa
\bqa
 \frac{d\hat \sigma(f \bar f \to Z\gamma) }{d \cos \vartheta^*}
&=&\frac{(\s-\mzd)}{2}~
\frac{d\hat \sigma(f \bar f \to Z\gamma) }{d \t}
\nonumber \\
&&
= \frac{(\s-\mzd)}{128\pi \s^2}
\sum_{\tau_1 \tau_2} \left [ |F^{\lam=1}_{\tau_1\tau_2}|^2
+|F^{\lam=-1}_{\tau_1\tau_2}|^2 \right ] ~. \label{dsig-ffZg}
\eqa
Note that in (\ref{dsig-ffZZ}), the identity of the two final $Z$
is taken into account by imposing the constrain $0 \leq
\cos\vartheta^* \leq 1$.\par

The Left-Right asymmetry measurable at an LC is defined by
\bqa
A_{LR}(e^- e^+ \to V_1V_2) &= &
[\frac{d\hat \sigma(e_L^- e_R^+ \to V_1V_2 ) }{d \cos \vartheta^*}
-\frac{d\hat \sigma(e_R^- e_L^+ \to V_1V_2 ) }{d \cos
\vartheta^*}]\times\nonumber\\
&&[\frac{d\hat \sigma(e_L^- e_R^+ \to V_1V_2 ) }
{d \cos \vartheta^*} +\frac{d\hat \sigma(e_R^- e_L^+ \to V_1V_2 ) }
{d \cos \vartheta^*}]^{-1} \nonumber \\
&=& \frac{\sum_{\tau_1 \tau_2} \left [ |F^{\lam=1}_{\tau_1\tau_2}|^2
-|F^{\lam=-1}_{\tau_1\tau_2}|^2 \right ]}
{\sum_{\tau_1 \tau_2} \left [ |F^{\lam=1}_{\tau_1\tau_2}|^2
+|F^{\lam=-1}_{\tau_1\tau_2}|^2 \right ]}
\label{ALR}
\eqa \par

Finally the $p_T$ distribution  at a hadron collider,
with c.m. energy $\sqrt{s}$, is determined by
\bq
\frac{d \sigma(H_1 H_2 \to V_1 V_2+... ) }{d p_T^2 dy_1dy_2}
= \tau \Sigma^{V_1V_2 }(x_a, x_b) ~ ~ , \label{dsig-pT}
\eq
where
\bqa
&& \Sigma^{ZZ}(x_a, x_b)=\sum_j
\left [ q_j(x_a)\bar q_j(x_b) + \bar q_j(x_a) q_j(x_b) \right ]
\frac{d\hat \sigma( q \bar q \to ZZ) }{d \t} ~, \label{Sxaxb-ZZ}
\\
&&  \Sigma^{Z\gamma }(x_a, x_b)=\sum_j
\left [ q_j(x_a)\bar q_j(x_b) + \bar q_j(x_a) q_j(x_b) \right ]
\frac{d\hat \sigma( q \bar q \to Z \gamma ) }{d \t} ~, \label{Sxaxb-Zg}
\eqa
The summation extend over all quark and antiquarks inside the hadrons
$H_1,~H_2$. Note that the distributions considered in this paper are
symmetrical in $\hat t$, $\hat u$ interchange.
$\tau =\s/s$, while $(x_a, x_b)$ are fully determined in terms of
the rapidities ($y_1,~y_2$) and the (opposite) transverse momenta
of the two final gauge bosons:

\bq
x_a={1\over2}[x_{1T}e^{y_1}+ x_{2T}e^{y_2}] \ \ \ \ \ \
x_b={1\over2}[x_{1T}e^{-y_1}+ x_{2T}e^{-y_2}]
\eq
\noindent
where
$x_{iT}={2\over \sqrt{s}}~(p^2_T+m^2_{V_i})^{{1\over2}}$. In the
integration over $y_1$ and $y_2$ we impose a cut at $|y_i|<2$.\par
We will also discuss the invariant mass and c.m. scattering angle
distributions given by
\bq
\frac{d \sigma(H_1 H_2 \to V_1 V_2+... ) }{d \hat s
d\cos\vartheta^*d\bar y}
= J\tau \Sigma^{V_1V_2 }(x_a, x_b) ~ ~ , \label{dsig-M}
\eq
where
$J(Z\gamma)={\hat s-m^2_Z\over2\hat s}$,
$J(ZZ)={1\over2}\sqrt{1-{4m^2_Z\over \hat s}}$.
$\bar y$ is the boost defined as $\bar y=y_1-y^*_1=y_2-y^*_2$,
with $thy^*_1=\beta_1^*\cos\vartheta^*$,
$thy^*_2=-\beta_2^*\cos\vartheta^*$
and $\beta^*_1={\hat s-m^2_Z\over \hat s^+m^2_Z}$, $\beta^*_2=1$ for
$Z\gamma$, but $\beta^*_1=\beta^*_2=\sqrt{1-{4m^2_Z\over\hat s}}$ for
$ZZ$. With these variables, $x_a=e^{\bar y}\sqrt{\tau}$,
$x_b=e^{-\bar y}\sqrt{\tau}$.

\clearpage

\clearpage

\clearpage

\clearpage

\clearpage

\clearpage

\clearpage

\clearpage

\clearpage

\end{document}